\documentclass[suppldata]{interact}

\usepackage[natbibapa,nodoi]{apacite}
\theoremstyle{plain}% Theorem-like structures provided by amsthm.sty

\theoremstyle{definition}

\theoremstyle{remark}

\usepackage[utf8]{inputenc}
\usepackage{amsmath}
\usepackage{mathrsfs}
\usepackage{amssymb}
\usepackage{color}
\usepackage{comment}
\usepackage{float}
\usepackage{graphicx}
\usepackage{bernsteinStyle}
\usepackage{float}
\usepackage{tikz}
\usepackage{booktabs}
\usepackage{pgfplots}
\usepackage{array}
\usepackage{bm}
\usepackage{makecell}
\newcolumntype{C}[1]{>{\centering\let\newline\\\arraybackslash\hspace{0pt}}m{#1}}
\newcommand\nhline[1]{\noalign{\hrule height #1}}
\usetikzlibrary{arrows,shapes,positioning}

\usetikzlibrary{shapes,arrows,calc,positioning}
\tikzstyle{bigblock} = [draw, fill=blue!20, rectangle, 
    minimum height=6em, minimum width=8em]
\tikzstyle{medblock} = [draw, fill=blue!20, rectangle, 
    minimum height=4em, minimum width=4em]    
\tikzstyle{mux} = [draw, fill=black!20, rectangle, 
    minimum height=5em, minimum width=0.1em]    
\tikzstyle{smallblock} = [draw, fill=blue!20, rectangle, 
    minimum height=3em, minimum width=4em]
\tikzstyle{sum} = [draw, fill=blue!20, circle, node distance=1cm]
\tikzstyle{signal} = [coordinate]
\tikzstyle{pinstyle} = [pin edge={to-,thin,black}]
\tikzstyle{block} = [draw, fill=blue!20, rectangle, 
    minimum height=3em, minimum width=6em]
\tikzstyle{blockS} = [draw, fill=blue!20, rectangle, 
    minimum height=3em, minimum width=4em]  
\tikzstyle{sum} = [draw, fill=blue!20, circle, node distance=1.5cm]
\tikzstyle{gain} = [draw, fill=blue!20, regular polygon, regular polygon sides = 3, node distance=1.25cm, shape border rotate = -90]
\tikzstyle{mult} = [draw, fill=blue!20, circle, node distance=1.25cm]
\tikzstyle{saturation block} = [draw,fill=blue!20, 
    	path picture={
    		\pgfpointdiff{\pgfpointanchor{path picture bounding box}{north east}}%
    		{\pgfpointanchor{path picture bounding box}{south west}}
    		\pgfgetlastxy\x\y
    		\tikzset{x=\x*.4, y=\y*.4}
    		\draw (-1,0) -- (1,0) (0,-1) -- (0,1); 
    		\draw (-1,-.7) -- (-.5,-.7) -- (.5,.7) -- (1,.7);
    	}]
\tikzstyle{sat atan} = [draw,fill=blue!20, 
        path picture={
          \pgfpointdiff{\pgfpointanchor{path picture bounding box}{north east}}%
            {\pgfpointanchor{path picture bounding box}{south west}}
          \pgfgetlastxy\x\y
          \tikzset{x=\x*.05, y=\y*0.3}
          \draw (-7.5,0) -- (7.5,0) (0,-1.2) -- (0,1.2); 
          \draw[scale = 1,domain=-7.5:7.5,smooth,variable=\x,black] plot ({\x},{(tanh((0.6*\x)))});
        }]
\tikzstyle{input} = [coordinate]
\tikzstyle{output} = [coordinate]

\pgfplotsset{compat=1.14}

\title{Identification of Self-Excited Systems Using Discrete-Time, Time-Delayed Lur'e Models}

\author{
\name{Juan A. Paredes\thanks{CONTACT Juan A. Paredes. Email: jparedes@umich.edu} and Dennis S. Bernstein\thanks{CONTACT Dennis S. Bernstein. Email: dsbaero@umich.edu}}
\affil{Department of Aerospace Engineering, University of Michigan, Ann Arbor, Michigan, USA}
}

\begin{document}

\maketitle

\begin{abstract}
This paper presents a system identification technique for systems whose output is asymptotically periodic under constant inputs. 
The model used for system identification is a discrete-time Lur'e model consisting of asymptotically stable linear dynamics, a time delay, a washout filter, and a static nonlinear feedback mapping.
For all sufficiently large scalings of the loop transfer function, these components cause divergence under small signal levels and decay under large signal amplitudes, thus producing an asymptotically oscillatory output.
A bias-generation mechanism is used to provide a bias in the oscillation.
The contribution of the paper is a least-squares technique that estimates the coefficients of the linear model as well as the parameterization of the continuous, piecewise-linear feedback mapping. 
\end{abstract}


\section{Introduction}

Nonlinear system identification is an exciting area of research with numerous challenges and open problems;  the  overview in \cite{SchoukensLjung} describes the status of the field and provides extensive references. 
The present paper focuses on nonlinear system identification for systems whose response to a constant input is asymptotically periodic;  a system of this type is called a  {\it self-excited system} (SES).
A classical example of a SES is the van der Pol oscillator, whose states converge to a limit cycle.
A SES, however, may have an arbitrary number of states and need not possess a limit cycle.
Overviews of SES are given in \cite{JENKINS2013167,Ding2010};
applications to chemical and biochemical systems are discussed in \cite{chance,gray_scott_1990,goldbeter_berridge_1996};
self-excited thermoacoustic oscillation  is discussed in \cite{Dowling1997,awad_1986,chen_2016};
and fluid-structure interaction and its role in aircraft wing flutter is discussed in \cite{blevins,friedmann,coller,cesnik}. 

A convenient model for SES is a feedback loop consisting of linear dynamics and a static nonlinear feedback mapping; a system of this type is called a {\it Lur'e system} \cite{khalil3rd}. 
Within the context of SES, Lur'e systems are considered in \cite{Ding2010,jian2004,Zanette_2017,Gusman_2016,CHATTERJEE20111860,gstan2007,Tomberg1989,chua1979,aguilar2009,hang2002,STAN2004LURELIN,savaresi2001}.
Self-oscillating discrete-time systems are considered in \cite{vrasvan98,amico2001,amico2004,amico2011}.

As discussed in \cite{DTLarxiv,DTLACC},
self-excited oscillations arise in Lur'e systems from a combination of stabilizing and destabilizing effects.
Destabilization at small signal levels causes the output to diverge, whereas stabilization at large signal levels causes the output to decay.

To provide a framework for SES system identification, this paper considers a discrete-time, time-delay Lur'e model consisting of an asymptotically stable linear system, a time delay, a washout filter, and a static nonlinear feedback mapping.
For all sufficiently large scalings of the loop transfer function, these components cause divergence under small signal levels and decay under large signal amplitudes, thus producing an asymptotically oscillatory output.
A bias-generation mechanism is used to provide a nonzero offset in the oscillation.
Similar features appear in \cite{jian2004,Zanette_2017,Gusman_2016,Ding2010,CHATTERJEE20111860}.
Conditions under which the Lur'e model used in the present paper is SES are given in \cite{DTLarxiv,DTLACC}.

%%%%%%%%%%%%%%%%%%%%%%%%%%%%%%%
The contribution of the present paper is the development of a technique for identifying SES using discrete-time, time-delayed Lur'e models.
In setting up the model structure, the user must choose the order of the linear discrete-time model and the number of steps delay.
Once these are chosen, the system identification method estimates the parameters of the linear discrete-time model as well as the static nonlinear feedback mapping, which is formulated as a continuous, piecewise-linear (CPL) function characterized by its slope in each interval of a user-chosen partition of the real line.
Although a nonlinear least-squares optimization technique can be used for parameter estimation, we adopt the approach of \cite{vanPelt2001,vanpelt}, which minimizes a bound on the least-square cost function that can be optimized by linear least squares.
%%%%%%%%%%%%%%%%%%%%%%%%%%%%%%%%%%%

The contents of the paper are as follows.
Section 2 introduces SES and the DTTDL model used for identification.
Section 3 describes the parameterization of the CPL functions used to approximate the nonlinear feedback mapping.
Section 4 presents the DTTDL/CPL model, which consists of the DTTDL model with the CPL mapping parameterized in Section 3.
Section 5 describes the least-squares technique for identifying SES using CTTDL/CPL, and Section 6 describes a variation of this technique for the constant-input case.
Section 7 presents numerical examples.

{\bf Notation.} 
$\BBR \isdef (-\infty,\infty),$
%
%$\BBZ\isdef\{\ldots, -1,0,1,\ldots\},$
%
$\mathbb{N}\isdef\{0,1,2,\ldots\}.$  
%
%$\BBP\isdef\{1,2,\ldots\}.$
%

%%%%%%%%%%%%%%%%%%%%%%%%%%%%%
\section{Identification of self-excited systems using discrete-time, time-delayed Lur'e models}
\begin{figure}[h!]
    \centering
    \resizebox{0.5\columnwidth}{!}{%
    \begin{tikzpicture}[>={stealth'}, line width = 0.25mm]
    \node [input, name=input] {};
    \node [smallblock, rounded corners, right = 0.75cm of input, minimum height = 0.6cm, minimum width = 0.8cm] (system) {$\SSS$};
    \node [output, right = 0.75cm of system] (output) {};
    \node [input, right = 2 cm of output] (inputM) {};
    \node [smallblock, rounded corners, right = 0.75cm of inputM, minimum height = 0.6cm, minimum width = 0.8cm] (systemM) {$\SM$};
    \node [output, right = 0.75cm of systemM] (outputM) {};
    \draw [draw,->] (input) -- node [name=u]{} node [very near start, above] {$v $} (system);
    \draw [->] (system) -- node [very near end, above] {$y$}(output);
    \draw [draw,->] (inputM) -- node [name=u]{} node [very near start, above] {$v$} (systemM);
    \draw [->] (systemM) -- node [very near end, above] {$y_\rmm$}(outputM);
    \end{tikzpicture}
    }
    \caption{Self-excited system $\SSS$ with input $v $ and   output $y,$ and model $\SM$ with input $v$ and output $y_\rmm.$   A system identification algorithm is used to construct a model $\SM$ that  captures the dynamics of $\SSS$.}
    \label{systemsFig}
\end{figure}
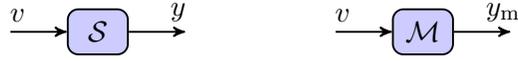

Let $\SSS$ be a discrete-time, self-excited system (SES) with input $v$ and output $y,$
and let $\SM$ be a discrete-time model with input $v$ and output $y_\rmm$
(see Figure \ref{systemsFig}).
The signals $v,y,y_\rmm$ are scalar.
The structure of $\SM$ is designed to capture the self-excited dynamics of $\SSS$ in the sense that, for all sufficiently large constant $v,$ 
there exist a nonconstant periodic function $\tau\colon{\mathbb N}\to\BBR$ and $k_0\in\BBN$ such that $\lim_{k\to\infty}|y_{k}-\tau_k| = 0$ and $\lim_{k\to\infty}|y_{\rmm,k}-\tau_{k + k_0}| = 0.$
In the case where $\SSS$ is a continuous-time SES, the output $y_k$ represents a sampled value of $y(t).$
In this paper, $\SM$ is chosen to be a discrete-time, time-delayed Lur'e model.
%%%%%%%%%%%%%%%%%%%%%%%%%%%%%%%%%%%%%

\begin{figure}[h!]
    \centering
    \resizebox{0.5\columnwidth}{!}{%
    \begin{tikzpicture}[>={stealth'}, line width = 0.25mm]
    \node [input, name=input] {};
    \node [smallblock, rounded corners, right of=input, minimum height=0.5cm, minimum width=0.5cm] (beta) {$\beta$};
    \node [sum, right = 0.5 cm of beta] (sum1) {};
    \node [draw = white] at (sum1.center) {$+$};
    \node [smallblock, rounded corners, right = 0.9cm of sum1, minimum height = 0.6cm, minimum width = 0.8cm] (system) {$G(\bfq)$};
    
    \draw [->] (sum1) -- node[name=usys, above] {$v_\rmb$} (system);
    \node [output, right = 2.3cm of system] (output) {};
    \node [smallblock, rounded corners, below = 0.6cm of system, minimum height = 0.6cm, minimum width = 0.8cm](diff){$G_\rmf(\bfq)$};
    \node [smallblock, rounded corners, right = 0.5cm of diff, minimum height = 0.6cm, minimum width = 0.8cm] (delay) {$G_d(\bfq)$};
    \node [smallblock, rounded corners, left = 0.5cm of diff, minimum height = 0.6cm, minimum width = 0.8cm](satq){$\SN$};
    \node [mult, below = 0.1cm of beta, minimum size=0.35cm] (mult1) {};
    \node [draw = white] at (mult1.center) {$\times$};
    
    \draw [draw,->] (input) -- node [name=u]{} node [very near start, above] {$v$} (beta);
    \draw [->] (u.center) |- (mult1);
    \draw [->] (beta) -- (sum1);
    \draw [->] (mult1) -| (sum1);
    \draw [->] (satq) -|  
    node [near start, above] {$v_\rmf$} (mult1);
    \draw [->] (system) -- node [name=y, very near end]{} node [very near end, above] {$y_\rmm$}(output);
    \draw [->] (y.center) |- (delay);
    \draw [->] (delay) -- node [above] {$y_\text{d}$} (diff);
    \draw [->] (diff) -- node [above] {$y_{\text{f}}$}(satq);
    \end{tikzpicture}
    }
    \caption{Discrete-time, time-delayed Lur'e  model with constant input $v$ and bias-generation mechanism.}
    \label{DT_TDL_offset_blk_diag}
\end{figure}
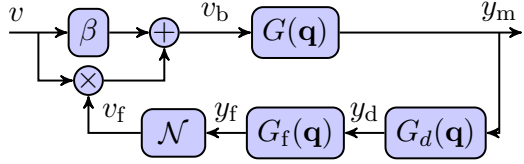

The discrete-time, time-delayed Lur'e (DTTDL) model shown in Figure \ref{DT_TDL_offset_blk_diag} incorporates the $n$th-order, asymptotically stable, strictly proper linear system 
\begin{equation}
    G(\bfq) = \frac{B(\bfq)}{A(\bfq)} = \frac{b_1 \bfq^{n-1} + \cdots + b_n}{\bfq^n + a_1 \bfq^{n-1} + \cdots + a_n}, \label{blockG}
\end{equation}
where $\bfq$ is the forward-shift operator,
the bias-generation mechanism
\begin{equation}
    v_\rmb = (\beta + v_\rmf)v, \label{biasGen}
\end{equation}
the time delay $G_\rmd(\bfq) = \bfq^{-d}$, where $d\ge0,$ the washout filter  
\begin{equation}
    G_\rmf(\bfq) = \frac{\bfq - 1}{\bfq}, \label{fbBlocks}
\end{equation}
and the nonlinear function $\SN\colon\BBR\to\BBR$
written as
\begin{equation}
    v_{\rmf,k} = \SN(y_{\rmf,k}). \label{nonL}
\end{equation}
Using $y_{\rmm,k} = G(\bfq) v_{\rmb,k},$ it follows that  
\begin{align}
A(\bfq)y_{\rmm,k} &= B(\bfq) v_{\rmb,k}\nn\\
&= B(\bfq) [\beta + \SN(y_{\rmf ,k})]v_{k},
\end{align}
and thus, for all $k\ge n+d+1,$ 
\begin{align}
    y_{\rmm,k}    
            &= (1 - A(\bfq))y_{\rmm,k} + B(\bfq)[\beta + \SN(y_{\rmf,k})]v_{k} \nn \\
            &= -a_1 y_{\rmm,k-1} - \cdots - a_n y_{\rmm,k-n}
            + \beta (b_1 v_{k-1} + \cdots + b_n v_{k-n}) \nn \\
            &\quad+ b_1 \SN(y_{\rmf,k-1})v_{k-1} + \cdots + b_n \SN(y_{\rmf,k-n})v_{k-n}, \label{CLeq1}
\end{align}
where
\begin{equation*}
    y_{\rmf,k}= y_{\rmm,k-d} - y_{\rmm,k-d-1}.
\end{equation*}
Note that the propagation of \eqref{CLeq1} depends on the initial output values $y_{\rmm,0},\ldots,y_{\rmm,n+d}.$ 

In \cite{DTLarxiv,DTLACC}, $\SN$ is assumed to be bounded, continuous, either nondecreasing or nonincreasing, and changes sign (positive to negative or vice versa) at zero; hence, $\SN(0) = 0$. 
Under these assumptions, it is shown in \cite{DTLarxiv,DTLACC} that, if the input $v$ is constant and sufficiently large, then the output $y_\rmm$ is nonconstant and asymptotically periodic.

\section{Parameterization of the continuous, piecewise-linear function $\SN$} \label{sec:sec4}

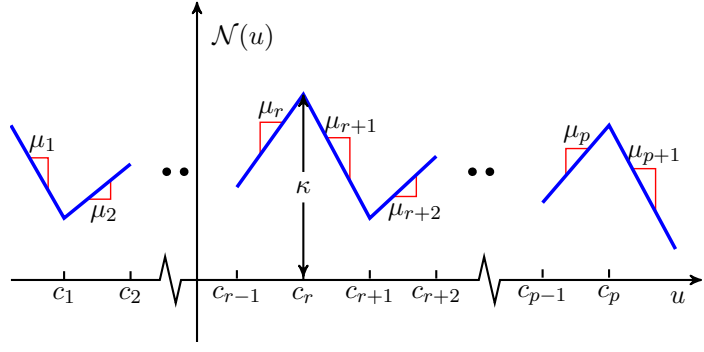
\begin{figure}[h!]
    \centering
    \resizebox{0.65\textwidth}{!}{%
    \begin{tikzpicture}[>={stealth'}]
        \begin{axis}
            [
            ticks=none,
            axis lines=left,
             axis line style={white},
            width=.8\textwidth,
            height=.45\textwidth, 
            xmin=-3.5,
            xmax=9.5,
            ymin=-1,
            ymax=10,
            xtick=\empty,
            ytick=\empty,
            ]
            %y axis Arrow
            \node (yAxisBase) at (axis cs:0,-2){};
            \node (yAxisTop) at (axis cs:0,10){};
            \node [below right = 0.05cm and 0.05cm of yAxisTop, xshift = -0.1cm](yAxisLabel){$\SN(u)$};
            %\node [below right = 0.05cm and 0.05cm of yAxisTop, xshift = -0.1cm](yAxisLabel){$g(u)$};
            \draw[line width = 0.3mm, ->](yAxisBase.center)--(yAxisTop.center);
            
            %x axis Arrow
            \node (xAxisBase) at (axis cs:-3.5,1){};
            \node (xAxisTop) at (axis cs:9.5,1){};
            \node [left =  0.5cm of xAxisTop.center, xshift = 0.4cm, yshift = -0.25cm](xAxisLabel){$u$};
            
            \node (xAxisP1) at (axis cs:-0.7,1){};
            \node (xAxisP2) at (axis cs:-0.6,1.75){};
            \node (xAxisP3) at (axis cs:-0.4,0.25){};
            \node (xAxisP4) at (axis cs:-0.3,1){};
            
            \node (xAxisP5) at (axis cs:5.3,1){};
            \node (xAxisP6) at (axis cs:5.4,1.75){};
            \node (xAxisP7) at (axis cs:5.6,0.25){};
            \node (xAxisP8) at (axis cs:5.7,1){};
            
            \draw[line width = 0.3mm, ->](xAxisBase.center)--(xAxisP1.center)--(xAxisP2.center)
            --(xAxisP3.center)--(xAxisP4.center)--(xAxisP5.center)--(xAxisP6.center)
            --(xAxisP7.center)--(xAxisP8.center)--(xAxisTop.center);
            
            %\draw[line width = 0.3mm, -](xAxisBase.center)--(xAxisP1.center);
            %\draw[line width = 0.3mm, -](xAxisP1.center)--(xAxisP2.center);
            %\draw[line width = 0.3mm, -](xAxisP2.center)--(xAxisP3.center);
            %\draw[line width = 0.3mm, -](xAxisP3.center)--(xAxisP4.center);
            %\draw[line width = 0.3mm, -](xAxisP4.center)--(xAxisP5.center);
            %\draw[line width = 0.3mm, -](xAxisP5.center)--(xAxisP6.center);
            %\draw[line width = 0.3mm, -](xAxisP6.center)--(xAxisP7.center);
            %\draw[line width = 0.3mm, -](xAxisP7.center)--(xAxisP8.center);
            %\draw[line width = 0.3mm, ->](xAxisP8.center)--(xAxisTop.center);
            
            %Triple dots
            %\node (dots1) at (axis cs:-0.5,4.5){ $\cdots$};
            %\node (dots2) at (axis cs:5.5,4.5){$\cdots$};
            
            \node (dots11) at (axis cs: -0.75,4.5){};
            \node (dots12) at (axis cs: 0.25,4.5){};
            \node (dots21) at (axis cs: 5,4.5){};
            \node (dots22) at (axis cs: 6,4.5){};
            
            \draw [line width=3.5pt, line cap=round, dash pattern=on 0pt off 2\pgflinewidth] (dots11) -- (dots12);
            \draw [line width=3.5pt, line cap=round, dash pattern=on 0pt off 2\pgflinewidth] (dots21) -- (dots22);
            
            %cr ticks and labels
            %c1 to c2
            \node (c1) at (axis cs:-2.5,0.5){$c_1$};
            \node [above = 1mm of c1](c1u){};
            \node [above = 0.1mm of c1, yshift = -3mm](c1d){};
            \draw [line width = 0.3mm, -](c1d)--(c1u);
            
            \node (c2) at (axis cs:-1.25,0.5){$c_2$};
            \node [above = 1mm of c2](c2u){};
            \node [above = 0.1mm of c2, yshift = -3mm](c2d){};
            \draw [line width = 0.3mm, -](c2d)--(c2u);
            
            %cr-1 to cr+2
            
            \node (cr_1) at (axis cs:0.75,0.5){$c_{r-1}$};
            \node [above = 1mm of cr_1](cr_1u){};
            \node [above = 0.1mm of cr_1, yshift = -3mm](cr_1d){};
            \draw [line width = 0.3mm, -](cr_1d)--(cr_1u);
            
            \node (cr) at (axis cs:2,0.5){$c_r$};
            \node [above = 1mm of cr](cru){};
            \node [above = 0.1mm of cr, yshift = -3mm](crd){};
            \draw [line width = 0.3mm, -](crd)--(cru);
            
            \node (cr1) at (axis cs:3.25,0.5){$c_{r+1}$};
            \node [above = 1mm of cr1](cr1u){};
            \node [above = 0.1mm of cr1, yshift = -3mm](cr1d){};
            \draw [line width = 0.3mm, -](cr1d)--(cr1u);
            
            \node (cr2) at (axis cs:4.5,0.5){$c_{r+2}$};
            \node [above = 1mm of cr2](cr2u){};
            \node [above = 0.1mm of cr2, yshift = -3mm](cr2d){};
            \draw [line width = 0.3mm, -](cr2d)--(cr2u);
            
            %cp-1 to cp
            
            \node (cp_1) at (axis cs:6.5,0.5){$c_{p-1}$};
            \node [above = 1mm of cp_1](cp_1u){};
            \node [above = 0.1mm of cp_1, yshift = -3mm](cp_1d){};
            \draw [line width = 0.3mm, -](cp_1d)--(cp_1u);
            
            \node (cp) at (axis cs:7.75,0.5){$c_p$};
            \node [above = 1mm of cp](cpu){};
            \node [above = 0.1mm of cp, yshift = -3mm](cpd){};
            \draw [line width = 0.3mm, -](cpd)--(cpu);
            
            %nodes for blue lines
            \node (pp0) at (axis cs:-3.5,6){};
            \node (pp1) at (axis cs:-2.5,3){};
            \node (pp2) at (axis cs:-1.25,4.75){};
            \node (ppr_1) at (axis cs:0.75,4){};
            \node (ppr) at (axis cs:2,7){};
            \node (ppr1) at (axis cs:3.25,3){};
            \node (ppr2) at (axis cs:4.5,5){};
            \node (ppp_1) at (axis cs:6.5,3.5){};
            \node (ppp) at (axis cs:7.75,6){};
            \node (ppp1) at (axis cs:9,2){};
            
            %mu triangles and labels
            \node (md011) at ($(pp0)!0.35!(pp1)$) {};
            \node (md012) at ($(pp0)!0.7!(pp1)$) {};
            \draw [red, line width = 0.2mm, -](md011.center)--(md012.center);
            \draw [red, line width = 0.2mm, -](md011.center)-|(md012.center);
            
            \node (md121) at ($(pp1)!0.35!(pp2)$) {};
            \node (md122) at ($(pp1)!0.7!(pp2)$) {};
            \draw [red, line width = 0.2mm, -](md121.center)--(md122.center);
            \draw [red, line width = 0.2mm, -](md121.center)-|(md122.center);
            
            \node (mdr_1r1) at ($(ppr_1)!0.35!(ppr)$) {};
            \node (mdr_1r2) at ($(ppr_1)!0.7!(ppr)$) {};
            \draw [red, line width = 0.2mm, -](mdr_1r1.center)--(mdr_1r2.center);
            \draw [red, line width = 0.2mm, -](mdr_1r1.center)|-(mdr_1r2.center);
            
            \node (mdrr11) at ($(ppr)!0.35!(ppr1)$) {};
            \node (mdrr12) at ($(ppr)!0.7!(ppr1)$) {};
            \draw [red, line width = 0.2mm, -](mdrr11.center)--(mdrr12.center);
            \draw [red, line width = 0.2mm, -](mdrr11.center)-|(mdrr12.center);
            
            \node (mdr1r21) at ($(ppr1)!0.35!(ppr2)$) {};
            \node (mdr1r22) at ($(ppr1)!0.7!(ppr2)$) {};
            \draw [red, line width = 0.2mm, -](mdr1r21.center)--(mdr1r22.center);
            \draw [red, line width = 0.2mm, -](mdr1r21.center)-|(mdr1r22.center);
            
            \node (mdp_1p1) at ($(ppp_1)!0.35!(ppp)$) {};
            \node (mdp_1p2) at ($(ppp_1)!0.7!(ppp)$) {};
            \draw [red, line width = 0.2mm, -](mdp_1p1.center)--(mdp_1p2.center);
            \draw [red, line width = 0.2mm, -](mdp_1p1.center)|-(mdp_1p2.center);
            
            \node (mdpp11) at ($(ppp)!0.35!(ppp1)$) {};
            \node (mdpp12) at ($(ppp)!0.7!(ppp1)$) {};
            \draw [red, line width = 0.2mm, -](mdpp11.center)--(mdpp12.center);
            \draw [red, line width = 0.2mm, -](mdpp11.center)-|(mdpp12.center);
            
            \node (mu1) at (axis cs:-2.9,5.4){$\mu_1$};
            \node (mu2) at (axis cs:-1.75,3.15){$\mu_2$};
            \node (mur) at (axis cs:1.4,6.5){$\mu_r$};
            \node (mur1) at (axis cs:2.9,6){$\mu_{r+1}$};
            \node (mur2) at (axis cs:4.1,3.2){$\mu_{r+2}$};
            \node (mup) at (axis cs:7.15,5.7){$\mu_p$};
            \node (mup1) at (axis cs:8.6,5.05){$\mu_{p+1}$};
            
            %blue lines
            
            \draw [blue, line width = 0.5mm, -](pp0.center)--(pp1.center)--(pp2.center);
            %\draw [blue, line width = 0.5mm, -](pp0.center)--(pp1.center);
            %\draw [blue, line width = 0.5mm, -](pp1.center)--(pp2.center);
            \draw [blue, line width = 0.5mm, -](ppr_1.center)--(ppr.center)--(ppr1.center)--(ppr2.center);
            %\draw [blue, line width = 0.5mm, -](ppr_1.center)--(ppr.center);
            %\draw [blue, line width = 0.5mm, -](ppr.center)--(ppr1.center);
            %\draw [blue, line width = 0.5mm, -](ppr1.center)--(ppr2.center);
            \draw [blue, line width = 0.5mm, -](ppp_1.center)--(ppp.center)--(ppp1.center);
            %\draw [blue, line width = 0.5mm, -](ppp_1.center)--(ppp.center);
            %\draw [blue, line width = 0.5mm, -](ppp.center)--(ppp1.center);
            
            %kappa line
            \node (kappaN) at (axis cs:2,4){$\kappa$};
            \node [above = 1.25mm of cr.center] (crN){};
            \draw [line width = 0.3mm, ->](kappaN.south)--(crN.center);
            \draw [line width = 0.3mm, ->](kappaN.north)--(ppr.center);
            
        \end{axis}
    \end{tikzpicture}
    }
    \caption{Parameterization of the CPL function $\SN.$  Note that $r$ and $\kappa$ fix $g$ along the ordinate axis.}
    \label{plnFunFig}
\end{figure}

In this section, we assume that $\SN$ is continuous and piecewise-linear (CPL), and we parameterize  $\SN$ as in \cite{vanPelt2001}.
Let $c_1 < \cdots < c_p$, let $(-\infty,c_1],(c_1,c_2],\ldots,(c_{p-1},c_p],(c_p,\infty)$ be a partition of the domain $\BBR$ of $\SN,$ and define the vector
\begin{equation}
    c \isdef {[ \arraycolsep=3.1pt\def\arraystretch{1.2} \begin{array}{ccc} c_1 & \cdots & c_p \end{array}]}^\rmT\in\BBR^p. \label{cEq}
\end{equation}
Furthermore, for all $i=1,\ldots,p+1,$ let $\mu_i$ denote the slope of $\SN$ in the $i$th partition interval, and define the slope vector
\begin{equation}
    \mu \isdef {[ \arraycolsep=3.1pt\def\arraystretch{1.2} \begin{array}{ccc} \mu_1 & \cdots & \mu_{p+1} \end{array}]}^\rmT\in\BBR^{p+1}. \label{muEq}
\end{equation}
Finally, letting $\kappa \in \BBR$ and  $r \in \{1, \ldots, p\},$  it follows that, for all $u \in \BBR,$
$\SN$ can be written as 
\begin{equation}
    \SN(u) = \mu^\rmT \eta(u) + \kappa, \label{PNL_Eq}
\end{equation}
where $\eta \colon \BBR \to \BBR^{p+1}$ is defined by
\begin{align}
    \eta (u) \isdef
    \begin{cases}
    \eta_{1} (u), & \delta(u) < r+1,\\
    \eta_{2} (u), & \delta(u) \geq r+1,
    \end{cases}
\end{align}
$\delta (u)\in\{1,\ldots,p+1\}$ is the index of the partition interval containing $u$, and
\begin{align}
    \eta_{1} (u) &\isdef {[ \arraycolsep=3.1pt\def\arraystretch{1.2} \begin{array}{cccccc} 0_{1 \times (\delta(u)-1)} & u - c_{\delta(u)} & c_{\delta(u)} - c_{\delta(u) + 1}
    & \cdots & c_{r-1} - c_{r} & 0_{1 \times (p + 1 - r)} \end{array}]}^\rmT ,\label{etac1Eq}\\
    \eta_{2} (u) &\isdef {[ \arraycolsep=3.1pt\def\arraystretch{1.2} \begin{array}{cccccc} 0_{1 \times r} & c_{r+1} - c_r & \cdots & c_{\delta(u) - 1} - c_{\delta(u) - 2}
    & u - c_{\delta(u) - 1} & 0_{1 \times (p + 1 - \delta(u))} \end{array}]}^\rmT .\label{etac2Eq}
\end{align}
Note that, if $\delta(u) = r,$ then $\eta_{1} (u) = {[ \arraycolsep=3.1pt\def\arraystretch{1.2} \begin{array}{cccccc} 0_{1 \times (\delta(u)-1)} & u - c_{\delta(u)} & 0_{1 \times (p + 1 - r)} \end{array}]}^\rmT ,$ 
whereas, if $\delta(u) = r + 1,$ then $\eta_{2} (u) \isdef {[ \arraycolsep=3.1pt\def\arraystretch{1.2} \begin{array}{cccccc} 0_{1 \times r} & u - c_{\delta(u) - 1} & 0_{1 \times (p + 1 - \delta(u))} \end{array}]}^\rmT .$
Since $\SN(c_r) = \kappa,$ it can be seen that $r$ and $\kappa$ fix $\SN$ along the ordinate axis, as shown in Figure \ref{plnFunFig}.
Hence, $\SN$ is parameterized by  $c,$ $\mu,$ $r,$ and $\kappa.$

\section{DTTDL model with a CPL nonlinear  feedback mapping}

In this section, we consider the DTTDL model in the case where $\SN$ is CPL; this is the DTTDL/CPL model.
In order to enforce $\SN(0) = 0$ (see Section 2), we let $\kappa = 0$ and assume that, for some $r,$ $c_r = 0.$
It thus follows from \eqref{CLeq1} and \eqref{PNL_Eq} that
\begin{align}
y_{\rmm,k}    
    &= -a_1 y_{\rmm,k-1} - \cdots - a_n y_{\rmm,k-n} \nn \\
    &\quad + \beta (b_1 v_{k-1} + \cdots + b_n v_{k-n}) \nn \\
    &\quad + b_1 \mu^\rmT \eta(y_{\rmf,k - 1})v_{k-1} + \cdots + b_n \mu^\rmT \eta(y_{\rmf, k-n})v_{k-n}. \label{CLeq_PNL3}
\end{align}
Now, defining
\begin{align}
    a \isdef {[ \arraycolsep=3.1pt\def\arraystretch{1.2} \begin{array}{ccc} a_1 & \cdots & a_n \end{array}]}^\rmT, \quad
    b \isdef {[ \arraycolsep=3.1pt\def\arraystretch{1.2} \begin{array}{ccc} b_1 & \cdots & b_n \end{array}]}^\rmT,
\end{align}
it follows that \eqref{CLeq_PNL3} can be written as
\begin{equation}
    y_{\rmm,k} = \phi_k^\rmT \theta, \label{CLeq_LS}
\end{equation}
where
\begin{equation}
    \theta \isdef \left[ \begin{array}{c} a \\ {\rm vec} (\mu b^\rmT) \\ \beta b \end{array} \right],\label{thetadefn}
\end{equation}
\begin{equation}
    \phi_k \isdef {[ \begin{array}{ccc} -\phi_{y,k}^\rmT & \phi_{\eta,k}^\rmT & \phi_{v,k}^\rmT \end{array} ]}^\rmT,
\end{equation}
and  
\begin{align}
    \phi_{y,k} &\isdef {[ \begin{array}{ccc} y_{\rmm,k-1} & \cdots & y_{\rmm,k-n} \end{array} ]}^\rmT, \label{phiykEq}\\
    \phi_{\eta,k} &\isdef {[ \begin{array}{ccc} v_{k-1} {\eta}^\rmT (y_{\rmf,k-1}) & \cdots & v_{k-n} {\eta}^\rmT (y_{\rmf,k-n}) \end{array} ]}^\rmT, \\
    \phi_{v,k} &\isdef {[ \begin{array}{ccc} v_{k-1} & \cdots & v_{k-n} \end{array} ]}^\rmT.
\end{align}

\section{Identification of DTTDL/CPL model parameters}\label{sec:LSCPL}

In this section, we present a least-squares identification technique for constructing a DTTDL/CPL model that approximates the response of the self-excited system $\SSS$.
Since we do not assume that $\SSS$ is a DTTDL system, the goal is to determine asymptotically stable $\hat G$ and CPL $\hat \SN$ such that the response of the identified model $\SM$ approximates the response of the true system $\SSS.$

The least-squares identification technique depends on choosing values of $n,d,c$;  these choices are denoted by $\hat{n},\hat{d},\hat c$.
In practice, $\hat{n},\hat{d},\hat c$ can be iteratively modified depending on the accuracy of the identification.
The goal is thus to obtain parameter estimates $\hat{a},\hat{b},\hat{\beta},\hat\mu$ for the DTTDL/CPL model.
In the special case where $\SSS$ is DTTDL or DTTDL/CPL, the parameters $\hat{a},\hat{b},\hat{\beta},\hat\mu$ can be viewed as estimates of $a,b,\beta,\mu$.

Next, let $l_\rmu \geq l_\rml \geq \hat{n} + \hat{d} + 1$ and, for all $k\in\{l_\rml - \hat{n} - \hat{d} - 1, \ldots, l_\rmu\},$ let $v_k$ and $y_k$ be the sampled measurements of $\SSS$ used for identification.
Then, define the least-squares cost
\begin{equation} \label{Jeq}
    J(\theta) \isdef  {||Y - \Phi \theta||}_2,
\end{equation}
where
\begin{equation}
    Y \isdef {[\begin{array}{ccc} y_{l_\rml} & \cdots & y_{l_\rmu} \end{array} ]}^\rmT, \label{LSeq}
\end{equation} 
and
\begin{equation}
    \Phi \isdef [\begin{array}{ccc} -\Phi_{Y} & \Phi_{\eta,Y} & \Phi_\rmV \end{array}], \label{PhiLSEq}
\end{equation}
where
\begin{equation}
    \Phi_{Y} \isdef \left[\def\arraystretch{1.2} \begin{array}{c} \phi_{Y,l_\rml}^\rmT \\ \vdots \\ \phi_{Y,l_\rmu}^\rmT \end{array}\right], \hspace{8mm} 
    \Phi_{\eta,Y} \isdef \left[\def\arraystretch{1.2} \begin{array}{c} \phi_{\eta, Y, l_\rml}^\rmT \\ \vdots \\ \phi_{\eta,Y,l_\rmu}^\rmT \end{array}\right], \hspace{8mm} 
    \Phi_{\rmV} \isdef \left[\def\arraystretch{1.2} \begin{array}{c} \phi_{v,l_\rml}^\rmT \\ \vdots \\ \phi_{v,l_\rmu}^\rmT \end{array}\right], 
\end{equation}
and
\begin{align}
    \phi_{Y,k} &\isdef {[ \begin{array}{ccc} y_{k-1} & \cdots & y_{k-\hat n} \end{array} ]}^\rmT, \\
    \phi_{\eta,Y,k} &\isdef {[ \begin{array}{ccc} v_{k-1} {\eta}^\rmT (y_{\rmf,Y,k-1}) & \cdots & v_{k-\hat n} {\eta}^\rmT (y_{\rmf,Y, k-\hat n}) \end{array} ]}^\rmT, \\
    y_{\rmf, Y,k} &\isdef y_{k- \hat d} - y_{k - \hat d - 1}. \label{PhiLSEqEnd}
\end{align}

Since $\theta$ given by \eqref{thetadefn} is not linear in $b,\mu,\beta,$ we derive an upper bound for $J(\theta),$ which is subsequently minimized.
To do this, let $\theta_\SA \in \BBR^{n(p+1)},$ define $\theta_\Lambda \isdef \beta b,$ and note that \eqref{LSeq} can be written as
\begin{align}
    J(\theta) &= {||Y - \Phi \theta + \Phi_{\eta,Y} \theta_\SA - \Phi_{\eta,Y} \theta_\SA||}_2 \nn \\
    &= {||Y + \Phi_{Y} a - \Phi_{\eta,Y} {\rm vec} (\mu b^\rmT) - \Phi_{\rmV} \theta_\Lambda +\Phi_{\eta,Y} \theta_\SA - \Phi_{\eta,Y} \theta_\SA||}_2 \nn \\
    &= {||Y - \Phi \tilde{\theta} + \Phi_{\eta,Y} (\theta_\SA - {\rm vec} (\mu b^\rmT)) ||}_2, \label{LSExteq}
\end{align}
where
\begin{equation}
    \tilde{\theta} \isdef \left[\def\arraystretch{1.2} \begin{array}{c} a \\ \theta_\SA \\ \theta_\Lambda \end{array}\right]. \label{theta_tilde_Eq}
\end{equation}
It follows from \eqref{LSExteq} that
%
%\label{Jineq}
\begin{align}
    J(\theta) &\leq {||Y - \Phi \tilde{\theta}||}_2 + {||\Phi_{\eta,Y} (\theta_\SA - {\rm vec} (\mu b^\rmT)) ||}_2 \nn \\
    &\leq {||Y - \Phi \tilde{\theta}||}_2 + \sigma_{\rm max} (\Phi_{\eta,Y}) {||\theta_\SA - {\rm vec} (\mu b^\rmT)||}_2  \nn  \\
    &= J_{\rm LS} (\tilde{\theta}) + \sigma_{\rm max} (\Phi_{\eta,Y}) J_\SA (\theta_\SA, \mu, b),\label{Jineq2}
\end{align}
where
\begin{gather}
    J_{\rm LS} (\tilde{\theta}) \isdef {||Y - \Phi \tilde{\theta}||}_2,\label{JLSdef}\\
    J_\SA (\theta_\SA, \mu, b) \isdef {||{\rm vec}^{-1}(\theta_\SA) - \mu b^\rmT||}_\rmF, \label{JSAdef}
\end{gather}
${||\cdot||}_\rmF$ denotes the Frobenius norm, and $\sigma_{\rm max}$ denotes the largest singular value.

The upper bound for $J(\theta)$ given by \eqref{Jineq2} is minimized by sequentially minimizing $J_{\rm LS}(\bar{\theta})$ and $J_\SA (\theta_\SA, \mu, b)$ to obtain
\begin{equation}
    \hat{\tilde{\theta}} \isdef \underset{\bar{\theta}_0 \in \BBR^{n(p+3)}}{\rm argmin}\, J_{\rm LS}(\bar{\theta}_0)= \left[ \begin{array}{c} \hat{a} \\ \hat{\theta}_\SA \\ \hat{\theta}_\Lambda \end{array} \right], \label{LSEsteq}
\end{equation}
where $\hat{\theta}_\SA \in \BBR^{n(p+1)}$ and  $\hat{\theta}_\Lambda \isdef \hat{\beta} \hat{b}.$
Note that $\hat{\tilde{\theta}}$ can be obtained by applying linear least-squares minimization to $J_{\rm LS}.$
Since $\hat{\beta}$ and $\hat{b}$ are unidentifiable from $\hat{\theta}_\Lambda,$
choosing an arbitrary nonzero value for $\hat{\beta}$ yields $\hat{b} = \hat{\theta}_\Lambda / \hat{\beta}.$

The following result is used to obtain $\hat{\mu} = \underset{\mu_0 \in \BBR^{p+1}}{\rm argmin}\, J_\SA (\hat{\theta}_\SA,\mu_0, \hat{b}).$

\begin{prop}\label{prop1}
Let $A \in \BBR^{n \times m}$, let $r \in \BBR^{m}$ be nonzero, and define $V \colon \BBR^n \to \BBR$ by
\begin{equation}\label{Veq}
    V(x) \isdef {||A - x r^\rmT||}_\rmF^2.
\end{equation}
\begin{equation}\label{argMinV}
    \underset{x\in \BBR^n}{\rm argmin}\, V(x) = (r^\rmT r)^{-1}{A r}.
\end{equation}
\end{prop}
\textbf{Proof.} For all $x\in\BBR^n,$  
\begin{equation}
    V(x)={\rm tr}(A^\rmT A) - 2 x^\rmT A r + x^\rmT x r^\rmT r,
\end{equation}
and thus
\begin{gather}
    V'(x) = -2 A r + 2 r^\rmT r x, \label{v2gradient}\\
     V''(x) = 2 r^\rmT r > 0. \label{v2hessian}
\end{gather}
It follows from \eqref{v2hessian} and \citep[Theorem 3.3.8, p.~115]{bazaraa2006} that $V$ is strictly convex,
which implies that $V$ has at most one minimizer.
Since
\begin{equation}\label{argMinVV}
    %\underset{x\in \BBR^n}{\rm argmin}\, ||V'(x)||_2 = \frac{A r}{r^\rmT r}.
     V'((r^\rmT r)^{-1}{A r}) = 0,
\end{equation}
\eqref{v2hessian} implies that $(r^\rmT r)^{-1}{A r}$ is a local minimizer of $V$.
Hence, \citep[Theorem 3.4.2, pp.~125, 126]{bazaraa2006} implies that 
$(r^\rmT r)^{-1}{A r}$ is the unique minimizer of $V.$
$\hfill \square$

\smallskip

Proposition \ref{prop1} implies that, for fixed $\hat{\theta}_\SA$ and $\hat{b},$ the value of $\hat\mu$ that minimizes $\hat\mu\mapsto J_\SA (\hat{\theta}_\SA,\hat\mu, \hat{b})$ is given by
\begin{equation}
    \hat{\mu} = \frac{{\rm vec}^{-1}(\hat{\theta}_\SA)\hat{b}}{\hat{b}^\rmT \hat{b}}.
\end{equation}
The identified DTTDL/CPL model $\SM$ is characterized by the  chosen parameters $\hat n, \hat d, \hat c,\hat \beta,$ as well as the estimated parameters $\hat a, \hat b,\hat \mu.$
Note that multiplying $\hat\beta$ by nonzero $\gamma \in \BBR$ results in the division of $\hat b$ by $\gamma$ and the multiplication of $\hat \mu$ by $\gamma$, which modifies the estimate of the nonlinear feedback mapping.
However, it follows from \eqref{CLeq_PNL3} that the response of the identified model remains unchanged.

\section{Identification of DTTDL/CPL model parameters with constant input}

This section considers a variation of the identification technique presented in the previous section for the case where $v$ is  constant, as typically occurs in self-excited systems.
For $v_k \equiv v_0$, \eqref{CLeq_PNL3} becomes
\begin{align}
    y_{\rmm,k}    &= -a_1 y_{\rmm,k-1} - \cdots - a_n y_{\rmm,k-n} + \beta v_0 (b_1 + \cdots + b_n)  \nn \\
    &\quad + v_0 [b_1 \mu^\rmT \eta(y_{\rmf,k - 1}) + \cdots + b_n \mu^\rmT \eta(y_{\rmf, k-n})]. \label{CLeq_PNL_v}
\end{align}
Then, \eqref{CLeq_PNL_v} can be expressed as \eqref{CLeq_LS}, where
\begin{gather}
    \theta \isdef \left[ \begin{array}{c} a \\ {\rm vec} (\mu b^\rmT) \\ \beta 1_{1 \times n} b \end{array} \right], \label{thetadefn2}\\
    \phi_{k} \isdef {[\begin{array}{ccc} -\phi_{y,k}^\rmT & \phi_{\eta,k}^\rmT & v_0 \end{array}]}^\rmT,
\end{gather}
$\phi_{y,k}$ is defined by \eqref{phiykEq}, and where
\begin{align}
    \phi_{\eta,k} &\isdef v_0 \ {[\begin{array}{ccc} {\eta}^\rmT (y_{\rmf,k-1}) & \cdots & {\eta}^\rmT (y_{\rmf,k-n}) \end{array}]}^\rmT.
\end{align}
Furthermore, $J(\theta)$ can be written as in \eqref{Jeq}, where $Y$ is defined by \eqref{LSeq} and $\Phi$ is defined by \eqref{PhiLSEq}--\eqref{PhiLSEqEnd}, where
\begin{align}
    \Phi_{\rmV} &\isdef v_0 1_{(l_\rmu - l_\rml + 1) \times 1}, \\
    \phi_{\eta,Y,k} &\isdef v_0 {[ \begin{array}{ccc} {\eta}^\rmT (y_{\rmf,Y,k-1}) & \cdots & {\eta}^\rmT (y_{\rmf,Y, k-\hat n}) \end{array} ]}^\rmT.
\end{align}

Since $\theta$ given by \eqref{thetadefn2} is not linear in $b,\mu,\beta,$ we derive an upper bound for $J(\theta),$ which is subsequently minimized.
Next, \eqref{Jeq} can be rewritten as in \eqref{LSExteq}, where $\theta_\SA \in \BBR^{n(p+1)},$ $\theta_\Lambda \isdef \beta 1_{1 \times n} b,$ and $\tilde{\theta}$ is defined as in \eqref{theta_tilde_Eq}.
Then, an upper bound for $J(\theta)$ can be derived as in $\eqref{Jineq2},$ where $J_{\rm LS}$ and $J_\SA$ are defined as in \eqref{JLSdef} and \eqref{JSAdef},  
and can be minimized by sequentially minimizing $J_{\rm LS}(\bar{\theta})$ and $J_\SA (\theta_\SA, \mu, b).$
Let $\hat{\theta}_\SA \in \BBR^{n(p+1)},$ define $\hat{\theta}_\Lambda \isdef \hat{\beta} 1_{1 \times n} \hat{b},$ and define $\hat{\bar{\theta}}$ as in \eqref{LSEsteq}.
Then $\hat{\bar{\theta}}$ can be obtained by minimizing $J_{\rm LS}.$

Next, \cite[Fact 11.16.39, p. 906]{bernstein2018} implies that, for fixed $\hat{\theta}_\SA,$ the rank-1 approximation of $\hat{\mu} \hat{b}^\rmT$ that minimizes $J_\SA (\hat{\theta}_\SA,\hat{\mu},\hat{b})$ is given by
\begin{equation} 
    \hat{\mu} \hat{b}^\rmT = \sigma_{\rm max}({\rm vec}^{-1}(\hat{\theta}_\SA))u_{\SA,1} v_{\SA,1}^\rmT, \label{mubMinEq}
\end{equation}
where $\sigma_{\rm max}$ denotes the largest singular value,
$u_{\SA,1}$ denotes the first left-singular vector of ${\rm vec}^{-1} (\hat{\theta}_\SA),$ and $v_{\SA,1}$ denotes the first right-singular vector of ${\rm vec}^{-1} (\hat{\theta}_\SA).$
Since $\hat\mu$ and $\hat b$ are unidentifiable from \eqref{mubMinEq}, choosing arbitrary nonzero $\beta_{\rm LS}\in\BBR$ and using it to separate \eqref{mubMinEq} yields
\begin{equation}
    \hat{\mu} = \beta_{\rm LS} \ \sigma_{\rm max} ({\rm vec}^{-1} (\hat{\theta}_\SA)) u_{\SA,1}, \hspace{5mm} \hat{b} = \frac{v_{\SA,1}}{\beta_{\rm LS}}.
\end{equation}
Finally, $\hat{\beta}$ is given by
\begin{equation}
    \hat{\beta} = \frac{\hat{\theta}_\Lambda}{1_{1 \times \hat{n}}  \hat{b}}.
\end{equation}

The identified DTTDL/CPL model $\SM$ is characterized by the chosen parameters $\hat n, \hat d, \beta_{\rm LS},$ and $\hat c,$ as well as the estimated parameters $\hat a, \hat b,$ $\hat \beta$ and $\hat \mu.$
Note that multiplying $\beta_{\rm LS}$ by nonzero $\gamma \in \BBR$ results in the division of $\hat b$ by $\gamma,$ and the multiplication of $\hat \mu$ (which scales $\hat\SN$) and $\hat{\beta}$ by $\gamma.$
However, it follows from \eqref{CLeq_PNL3} that the response of the identified model remains unchanged.

\section{Numerical examples}
In this section, we present numerical examples to illustrate identification of DTTDL/CPL models.
Recursive least squares (RLS) is used for regression, as presented in \cite{astrom,RLS2019}.
Table \ref{tab_Examples} summarizes the details of the numerical examples considered in this section.
The identified systems include three DTTDL systems, one continuous-time, time-delayed Lur'e (CTTDL) system, the Van der Pol (VdP) system with output bias, and the predator-prey Lotka-Volterra system. For continuous-time systems, $T_\rmd$ is the time delay.
For examples \ref{ex1} to \ref{ex4}, it is assumed that the input $v$ is known.
However, since examples \ref{ex5} and \ref{ex6} do not involve an external input, an arbitrary value of the input $v$ is used to facilitate identification of the DTTDL model.
Note that the outputs of the systems in Example 7.2 and Example 7.3 are asymptotically periodic under sufficiently large constant inputs, despite the fact that the nonlinearities in these systems do not satisfy the assumptions on $\SN$ in Section 2.

\begin{table}[h]
\vspace{0.7 em}
\caption{Examples for DTTDL/CPL System Identification}
\label{tab_Examples}
\centering
\setlength{\tabcolsep}{2.5pt}
\setlength{\arrayrulewidth}{0.2mm}
\renewcommand{\arraystretch}{1.5}
\begin{tabular}{!{\vrule width 0.6mm} c !{\vrule width 0.6mm} c !{\vrule width 0.6mm} c !{\vrule width 0.6mm} c !{\vrule width 0.6mm} c !{\vrule width 0.6mm} C{8.2em} !{\vrule width 0.6mm} C{8em} !{\vrule width 0.6mm}}
\nhline{0.6mm}
\bf Example & \bf System Type & $\bm n$ & $\bm d$ & $T_{\rmd}$ & $\bm \SN$ & \bf Parameters\\
\nhline{0.6mm}
\ref{ex1} & DTTDL & 2 & 4 & n/a & CPL, monotonic, odd  & $\beta = 7.5$ \\
\hline
%\addlinespace[0.2cm]
\ref{ex2} & DTTDL & 3 & 4 & n/a & C$^\infty,$   monotonic, not odd & $\beta = 5$ \\
\hline
%\addlinespace[0.2cm]
\ref{ex3} & DTTDL & 6 & 0 & n/a & C$^\infty,$ not monotonic, odd  & $\beta = 2.5$ \\
\hline
%\addlinespace[0.2cm]
\ref{ex4} & CTTDL & 2 & n/a & 0.1 s & C$^\infty,$ monotonic, odd  & $\beta = 50,$ $T_\rms = 0.1$ s \\
\hline
%\addlinespace[0.2cm]
\ref{ex5} & VdP w/bias & 2 & n/a & n/a & C$^\infty,$ multivariable & $\mu_0 = 1,$ $\overline{y} = 10,$ $T_\rms = 0.1$ s \\
\hline
%\addlinespace[0.2cm]
\ref{ex6} & Lotka-Volterra & n/a & n/a & n/a & n/a & \makecell{$\zeta = 2/3, \varrho = 4/3,$ \\ $\xi = 1, \varphi = 1,$ \\ $T_\rms = 0.1$ s}\\
\nhline{0.6mm}
\end{tabular}
%\vspace{-1em}
\end{table}% \end{table}
% %
\clearpage
% %
\refstepcounter{subsection}
\subsection*{Example \ref{ex1}: DTTDL system with CPL, monotonic, odd $\SN$} \label{ex1}
\begin{figure}[h]
    \centering
    \includegraphics[width=0.85\textwidth]{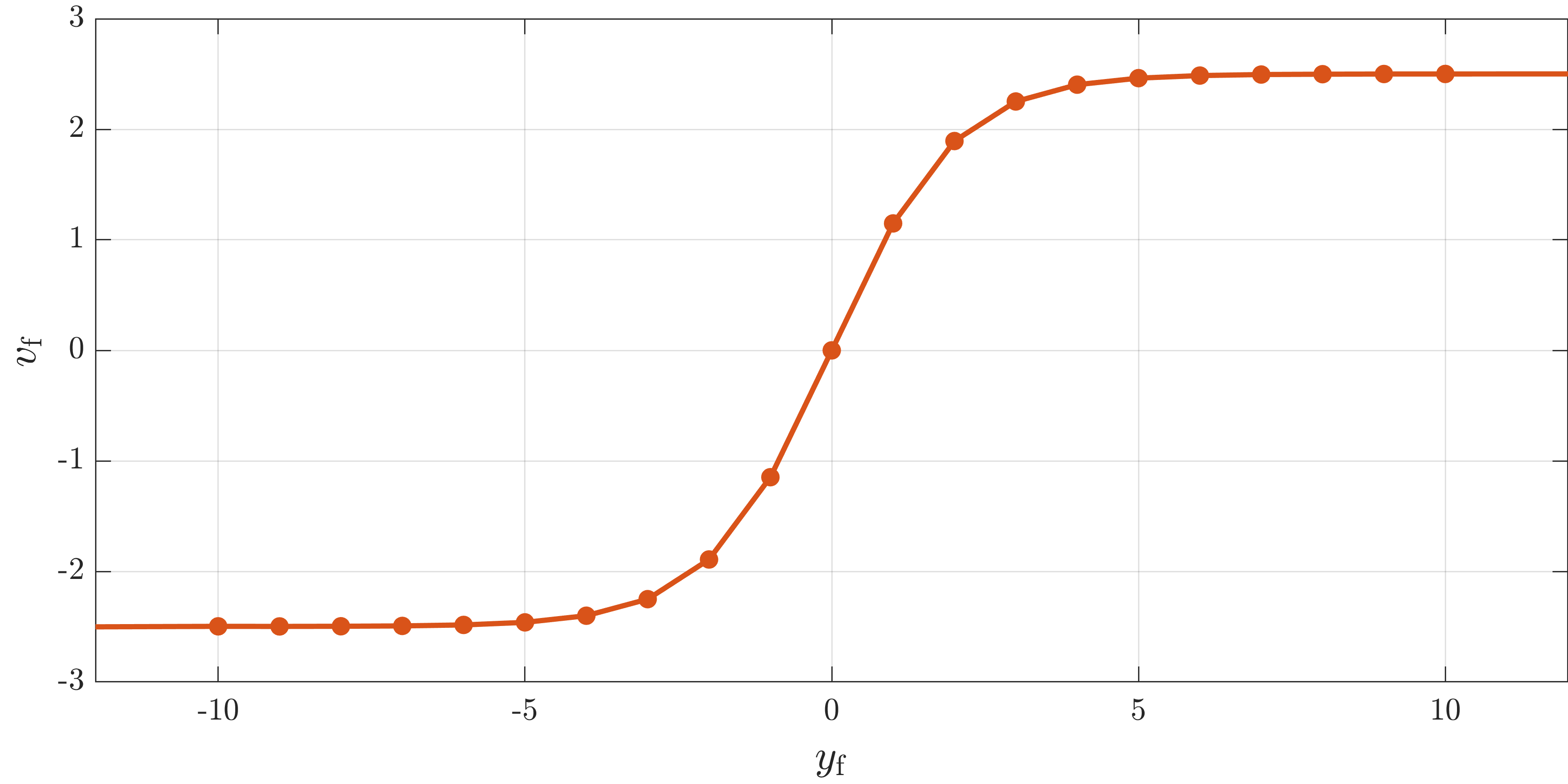}
    \caption{Example \ref{ex1}:  Piecewise-linear feedback mapping $\SN$.}
    \label{ex1_NL_og}
\end{figure}

Consider the DTTDL system $\SSS$ with $\beta = 7.5,$ $d = 4,$ 
\begin{equation} \label{ex1_eq}
    %G(\bfq) = \frac{\bfq - 0.5}{(\bfq - 0.8 + \jmath 0.4)(\bfq - 0.8 - \jmath 0.4)},
    G(\bfq) = \frac{\bfq - 0.5}{\bfq^2 - 1.6 \bfq + 0.8},
\end{equation}
and the CPL, monotonic, odd feedback mapping $\SN$ shown in Figure \ref{ex1_NL_og}. 
The domain  of $\SN$ is partitioned by
$
    c = {[\begin{array}{ccccccc} -10 & -9 & \cdots & 9 & 10 \end{array}]}^\rmT,
$
and $\SN$ is constructed such that, for all $i\in\{1,\ldots,21\},$ $\SN(c_i) = 2.5 \tanh(1.2 c_i / 2.5).$
To obtain data for identification, $y_0,\ldots,y_6$ are generated randomly, and, for all $k\ge0,$ $v_k$ is a gaussian random variable with mean $5$ and standard deviation $\sqrt{1.5}.$
For all $k\ge7,$ $y_k$ is generated by simulating $\SSS$ with \eqref{ex1_eq}.  The same technique is used in all subsequent examples.

For least-squares identification of the DTTDL/CPL model parameters, we let $\hat{c} = c$ and $\hat{\beta} = \beta$, and we apply RLS with $\theta_0 = 0,$ $P_0 = {10}^{6}$ and $\lambda = 1$ using data in $[100,25000].$ 
The standard deviation of the sensor noise 
is chosen to be $\sqrt{1.5},$ which yields a measurement signal-to-noise ratio (SNR) of approximately $40$ dB.

\begin{figure}[h]
    \centering
    \includegraphics[width=0.7\textwidth]{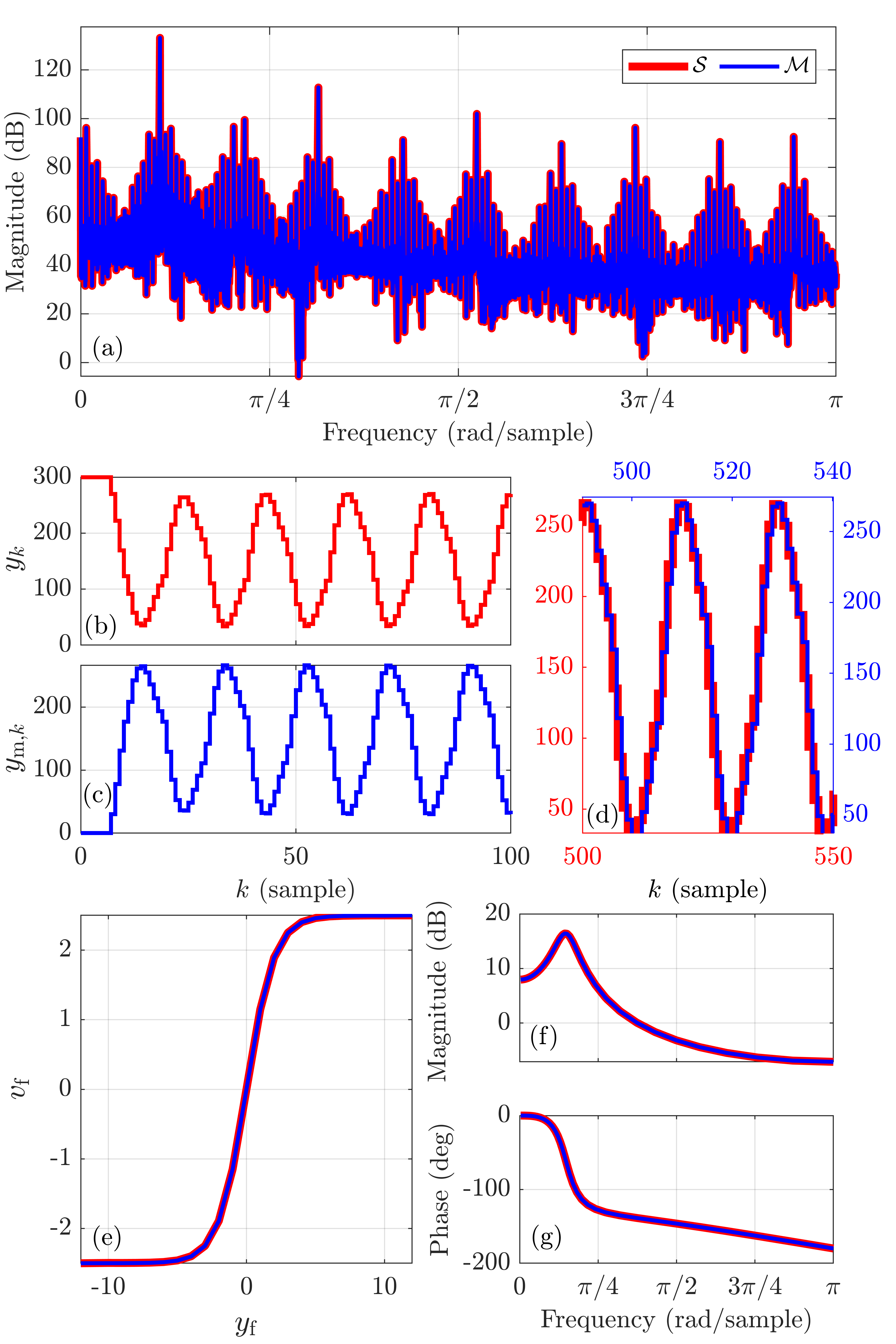}
    \caption{Example \ref{ex1}: Least-squares identification of DTTDL/CPL model parameters using  noiseless measurements with $\hat{n} = n$ and $\hat{d} = d.$ (a) compares the power spectral density (PSD) of the output of $\SM$ with the  PSD of the output of $\SSS$. (b) shows the output of $\SSS$ with $v_k \equiv 8$ and with $y_k = 300$ for all $k \in [0,6]$. (c) shows the output of $\SM$ with $v_k \equiv 8$ and with $y_{\rmm,k} = 0$ for all $k \in [0,6]$. (d) shows the output of $\SSS$ on $[500,550]$ and the output of $\SM$ on $[491,540].$ (e) shows the true and estimated nonlinearities. (f) and (g) show the frequency responses of the linear dynamics of $\SSS$ and $\SM$.}
    \label{ex1_ID_noiseless}
\end{figure}

\begin{figure}[h]
    \centering
    \includegraphics[width=0.7\textwidth]{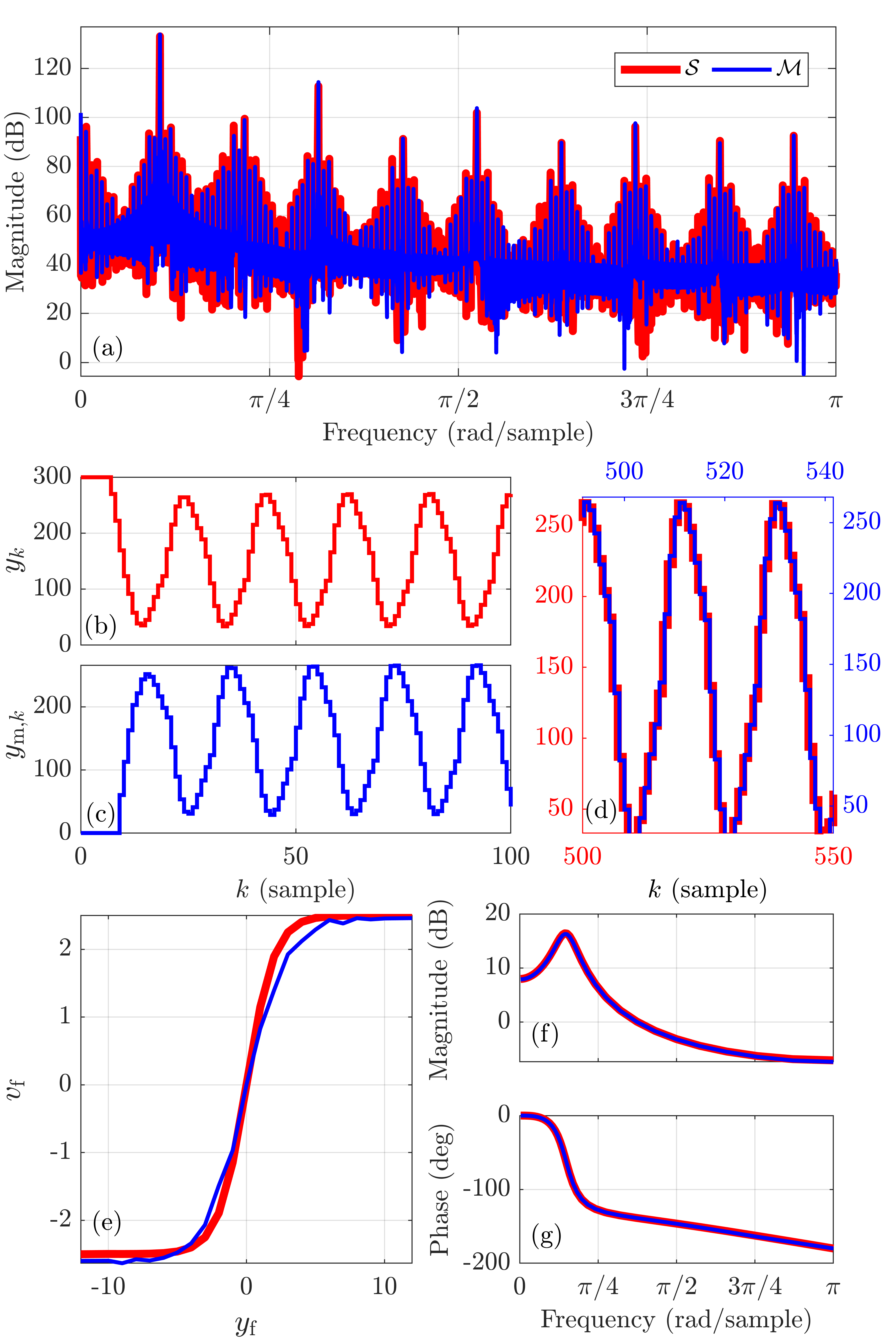}
    \caption{Example \ref{ex1}: Least-squares identification of DTTDL/CPL model parameters using  noisy measurements for $\hat{n} = 4$ and $\hat{d} = d.$ (a) compares the PSD of the output of $\SM$ with the PSD of the output of  $\SSS$. (b) shows the output of $\SSS$ with $v_k \equiv 8$ and with $y_k = 300$ for all $k \in [0,6]$. (c) shows the output of $\SM$ with $v_k \equiv 8$ and with $y_{\rmm,k} = 0$ for all $k \in [0,6]$. (d) shows the output of $\SSS$ on $[500,550]$ and the output of $\SM$ on $[492,541]$. (e) shows  the true and  estimated nonlinearities. (f) and (g) show the frequency responses of the linear dynamics of $\SSS$ and $\SM$.}
    \label{ex1_ID_noise}
\end{figure}

\begin{figure}[h]
    \centering
    \includegraphics[width=0.7\textwidth]{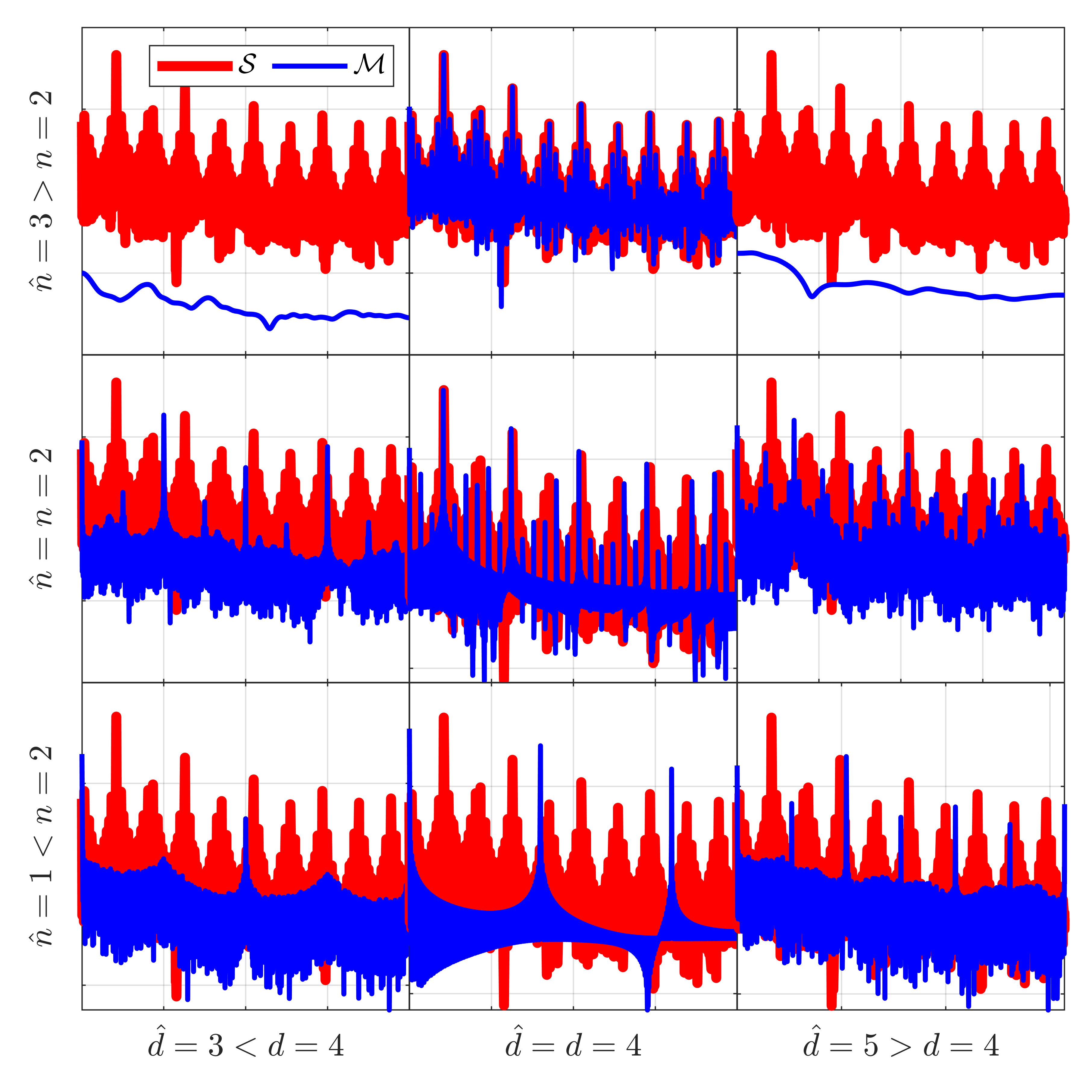}
    \caption{Example \ref{ex1}: For $\hat{n}\in\{1,2,3\}$ and $\hat{d}\in\{3,4,5\}$, these plots compare the PSD of the output of $\SM$ identified using noisy measurements with the PSD of the output of $\SSS$.}
    \label{ex1_ID_grid}
\end{figure}

To assess the accuracy of the  identified model, the input $v_k \equiv 8$ is applied to the system $\SSS$ with the initial conditions $y_k = 300$ for all $k\in[0,6],$ as well as the identified model $\SM$ with the initial conditions $y_{\rmm,k} = 0$ for all $k\in[0,6].$
The response of the identified model based on  noiseless measurements with $\hat{n} = n$ and $\hat{d} = d$ is shown in Figure \ref{ex1_ID_noiseless}, 
and the response of the identified model based on  noisy measurements with $\hat{n} = 4$ and $\hat{d} = d$ is shown in Figure \ref{ex1_ID_noise}. 
Figure \ref{ex1_ID_grid} compares the power spectral density (PSD) of the output of $\SM$ for $\hat{n} \in \{1, 2, 3\}$ and $\hat{d} \in \{3, 4, 5\}$ obtained using noisy measurements with the PSD of the output of $\SSS$.

\clearpage
\refstepcounter{subsection}
\subsection*{Example \ref{ex2}: DTTDL system with $C^\infty,$ monotonic, not odd $\SN$} \label{ex2}
Consider the DTTDL system $\SSS$ with $\beta = 5,$ $d = 4,$ 
\begin{equation}\label{ex2_eq}
    G(\bfq) = \frac{\bfq^2 -2.3 \bfq - 1.5725}{\bfq^3 - 2.35 \bfq^2 + 2 \bfq - 0.6},
\end{equation}
and $\SN(y_\rmf) = 2.5 \tanh \left(1.2 (y_\rmf - 3)/2.5\right) + 2.2342,$ which is monotonic but not odd.
To obtain data for identification, $y_0, \ldots, y_7$ are generated randomly, and, for all $k\geq 0,$ $v_k$ is a gaussian random variable with mean 4 and standard deviation $\sqrt{2}.$
For all $k\geq 8,$ $y_k$ is generated by simulating $\SSS$ with \eqref{ex2_eq}. 
For least-squares identification of the DTTDL/CPL model parameters, we let 
$
    \hat{c} = {[\begin{array}{ccccccc} -10 & -9 & \ldots & 9 & 10 \end{array}]}^\rmT
$
and $\hat{\beta} = \beta,$ and we apply RLS with $\theta_0 = 0,$ $P_0 = {10}^{6}$ and $\lambda = 1$ using data in $[100, 25000].$
The standard deviation of the sensor noise is chosen to be $\sqrt{2.65}$, which yields a measurement SNR of approximately $40$ dB.

\begin{figure}[h]
    \centering
    \includegraphics[width=0.7\textwidth]{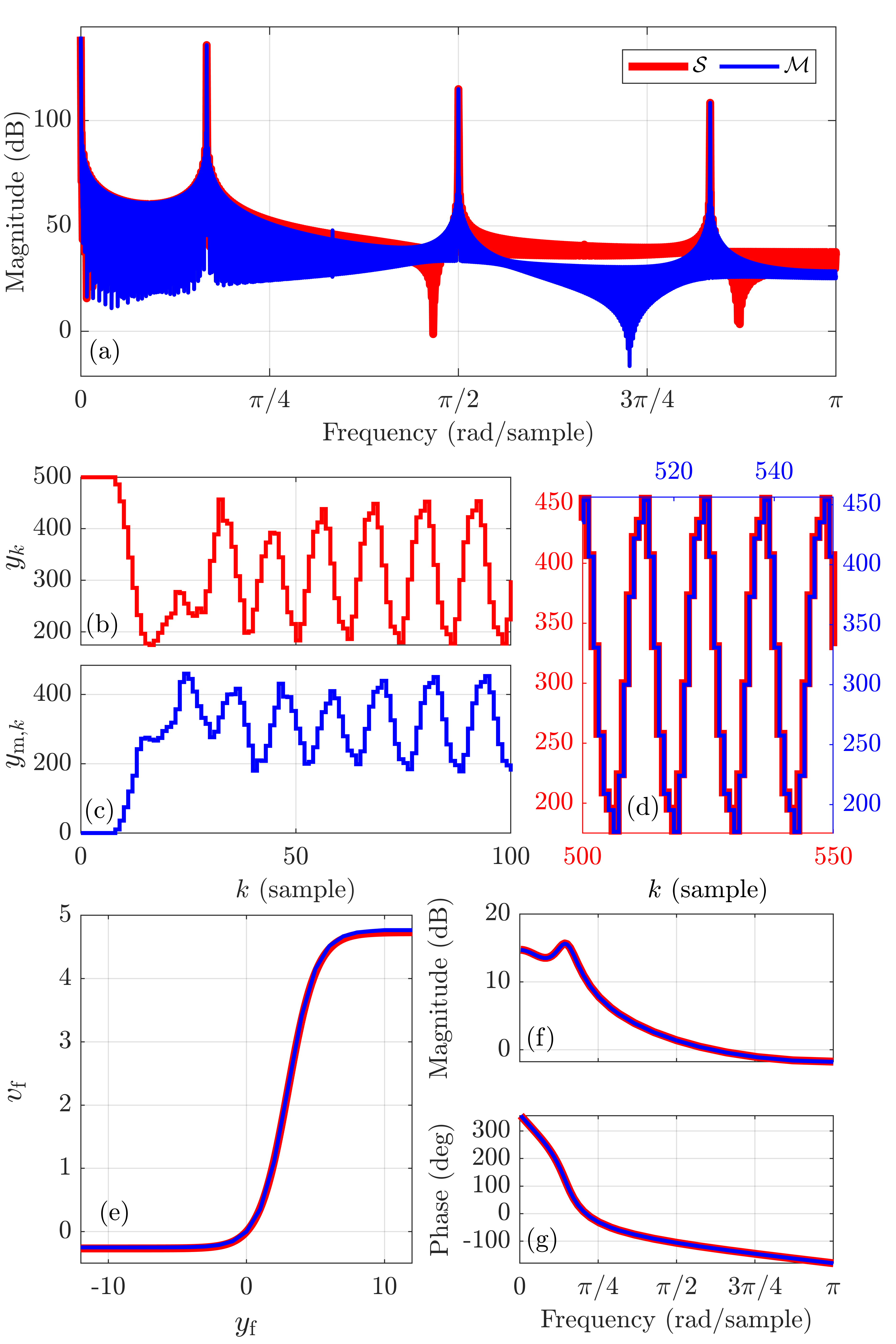}
    \caption{Example \ref{ex2}: Least-squares identification of DTTDL/CPL model parameters using  noiseless measurements with $\hat{n} = n$ and $\hat{d} = d.$ (a) compares the PSD of the output of $\SM$ with the PSD of the output of  $\SSS$. (b) shows the output of $\SSS$ with $v_k \equiv 8$ and with $y_k = 500$ for all $k \in [0,7]$. (c) shows the output of $\SM$ with $v_k \equiv 8$ and with $y_{\rmm,k} = 0$ for all $k \in [0,7]$. (d) shows the output of $\SSS$ on $[500,550]$ and the output of $\SM$ on $[502,552].$ (e) shows  the true and estimated nonlinearities. (f) and (g) show the frequency responses of the linear dynamics of $\SSS$ and $\SM$.}
    \label{ex2_ID_noiseless}
\end{figure}

\begin{figure}[h]
    \centering
    \includegraphics[width=0.7\textwidth]{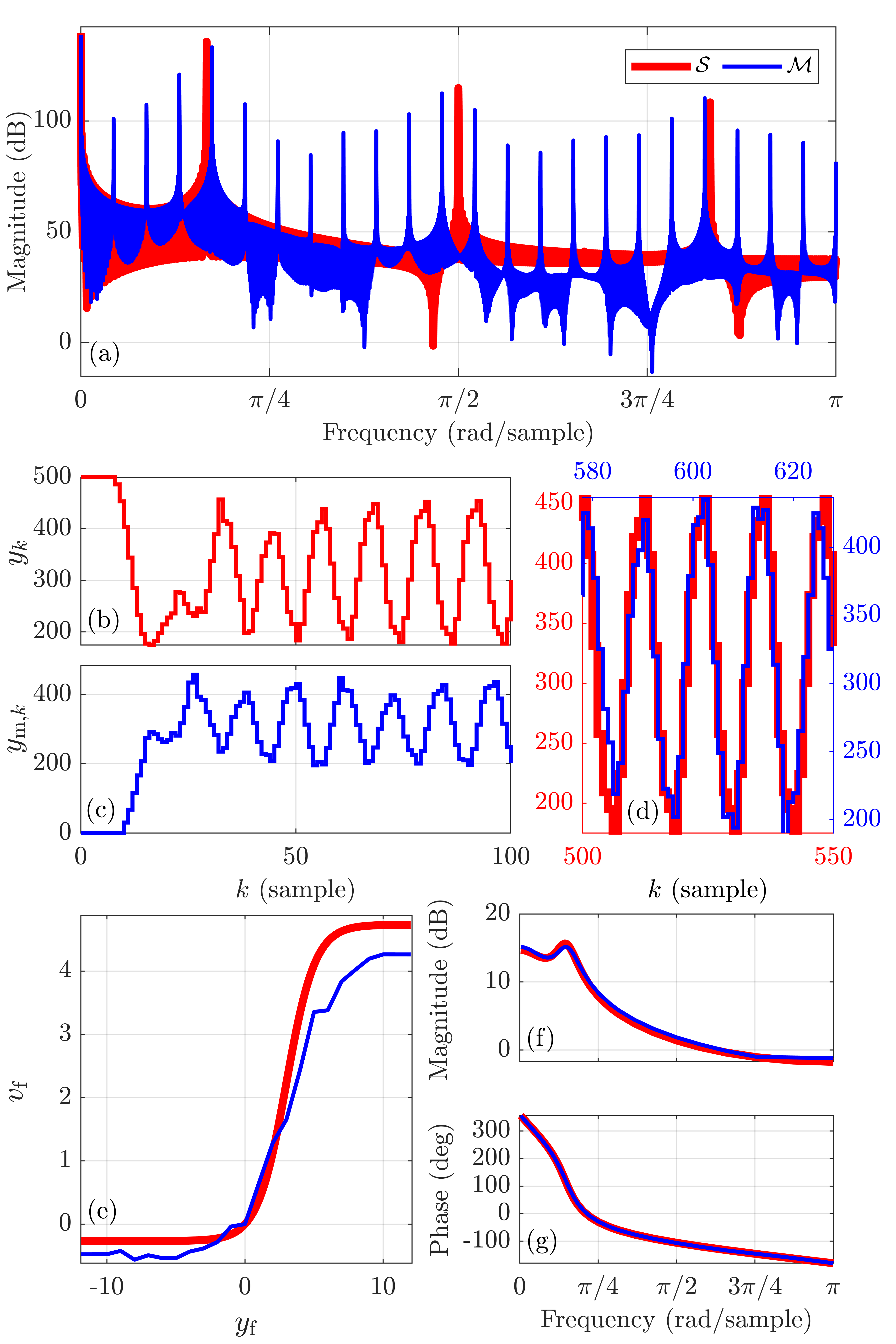}
    \caption{Example \ref{ex2}: Least-squares identification of DTTDL/CPL model parameters using  noisy measurements for $\hat{n} = 5$ and $\hat{d} = d.$ (a) compares the PSD of the output of $\SM$ with the PSD of the output of  $\SSS$. (b) shows the output of $\SSS$ with $v_k \equiv 8$ and with $y_k = 500$ for all $k \in [0,7]$. (c) shows the output of $\SM$ with $v_k \equiv 8$ and with $y_{\rmm,k} = 0$ for all $k \in [0,7]$. (d) shows the output of $\SSS$ on $[500,550]$ and the output of $\SM$ on $[579,629].$ (e) shows  the true and estimated nonlinearities. (f) and (g) show the frequency responses of the linear dynamics of $\SSS$ and $\SM$.}
    \label{ex2_ID_noise}
\end{figure}

\begin{figure}[h]
    \centering
    \includegraphics[width=0.7\textwidth]{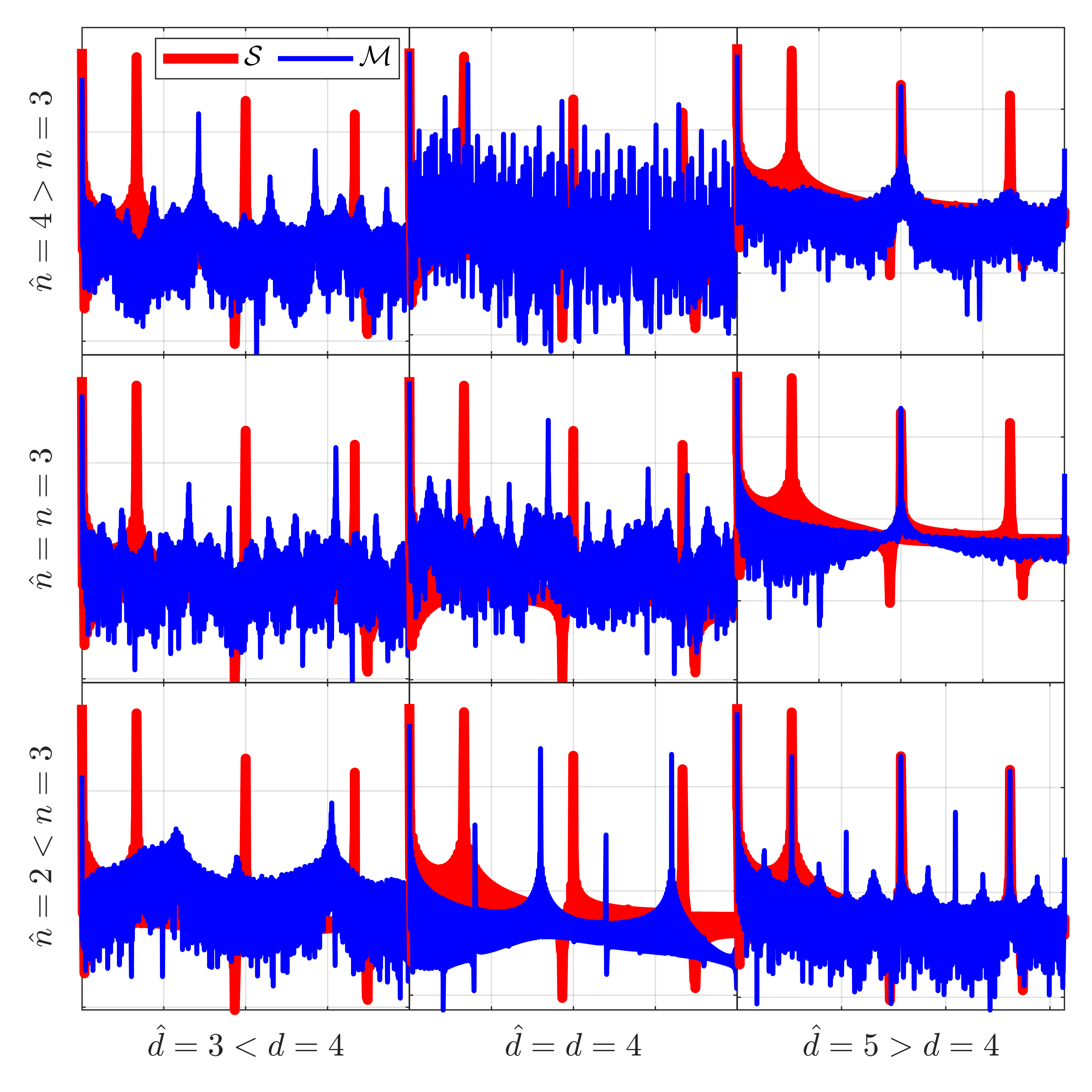}
    \caption{Example \ref{ex2}: For $\hat{n}\in\{2,3,4\}$ and $\hat{d}\in\{3,4,5\}$,  these plots compare the PSD of the output of $\SM$ identified using noisy measurements with  the PSD of the output of $\SSS$.}
    \label{ex2_ID_grid}
\end{figure}

To assess the accuracy of the identified model, the input $v_k \equiv 8$ is applied to the system $\SSS$ with the initial conditions $y_k = 500$ for all $k\in[0,7],$ as well as the identified model $\SM$ with the initial conditions $y_{\rmm,k} = 0$ for all $k\in[0,7].$
The response of the identified model based on  noiseless measurements with $\hat{n} = n$ and $\hat{d} = d$ is shown in Figure \ref{ex2_ID_noiseless},
and the response of the identified model based on  noisy measurements with $\hat{n} = 5$ and $\hat{d} = d$ is shown in Figure \ref{ex2_ID_noise}. 
Figure \ref{ex2_ID_grid} compares the PSD of the output of $\SM$ for $\hat{n} \in \{2, 3, 4\}$ and $\hat{d} \in \{3, 4, 5\}$ obtained using noisy measurements with the PSD of the output of $\SSS$.

\clearpage
\refstepcounter{subsection}
\subsection*{Example \ref{ex3}: DTTDL system with $C^\infty,$ not monotonic, odd $\SN$} \label{ex3}
Consider the DTTDL system $\SSS$ with $\beta = 1,$ $d = 0,$ 
\begin{equation}
    G(\bfq) = \frac{\bfq^2 + 1.5 \bfq + 0.8125}{\bfq^6 - 3.5442 \bfq^5 + 5.21974 \bfq^4 - 3.92160 \bfq^3 + 1.5316 \bfq^2 - 0.2722 \bfq - 0.02153},
\end{equation}
and
\begin{equation}\label{ex3_eq}
   \SN (y_\rmf) = - \SN_{\rm max} \frac{1}{\sigma_\SN \sqrt{2 \pi}} e^{-\half {((y_\rmf + \mu_\SN) / \sigma_\SN)}^2}
    + \SN_{\rm max} \frac{1}{\sigma_\SN \sqrt{2 \pi}} e^{-\half {((y_\rmf - \mu_\SN) / \sigma_\SN)}^2},
\end{equation}
which is not monotonic and odd, with $\SN_{\rm max} = 4,$ $\sigma_\SN = 1.75,$ and $\mu_\SN = 4.$

\begin{figure}[h]
    \centering
    \includegraphics[width=0.7\textwidth]{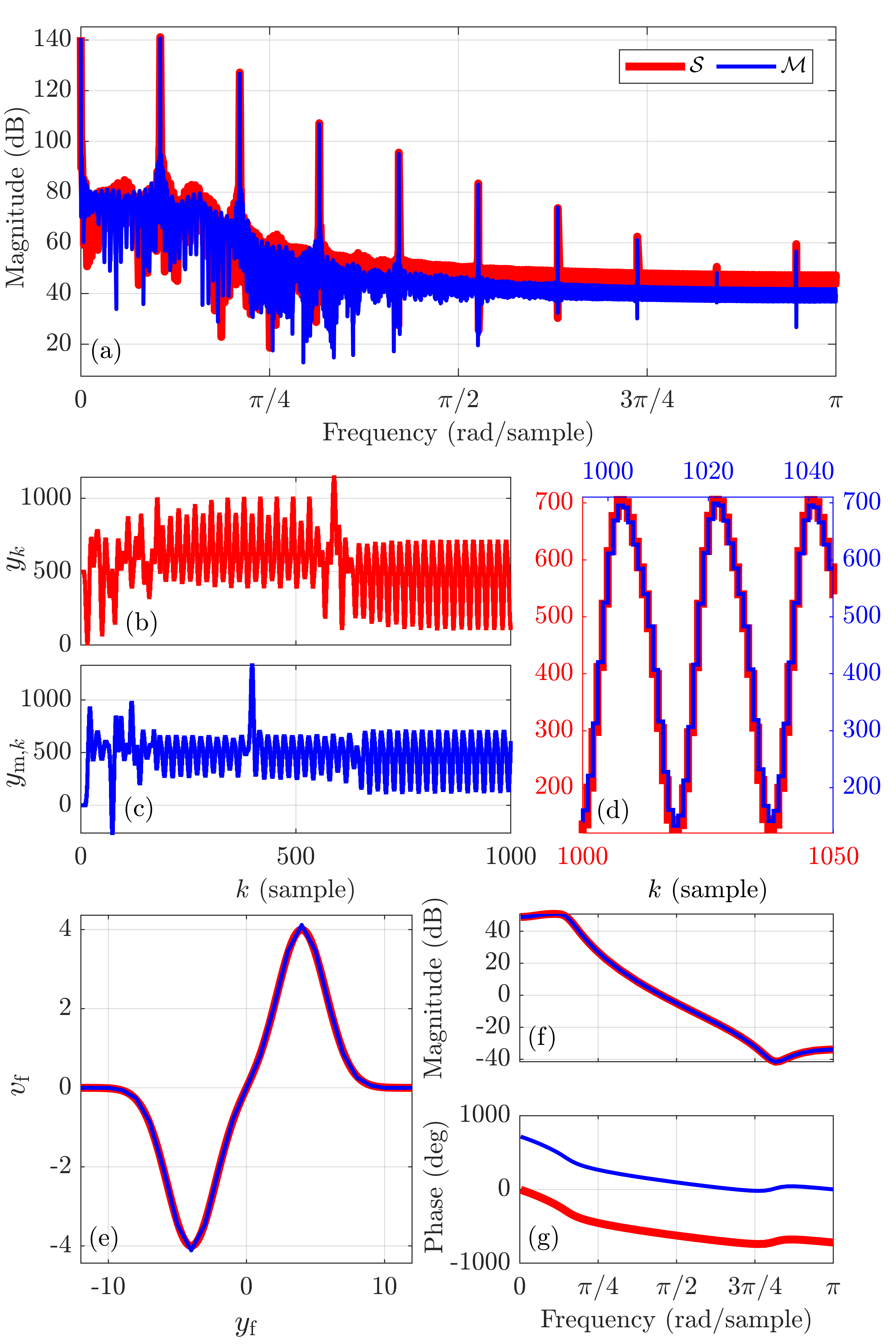}
    \caption{Example \ref{ex3}: Least-squares identification of DTTDL/CPL model parameters using noiseless measurements with $\hat{n} = n$ and $\hat{d} = d.$ (a) compares the PSD of the output of $\SM$ with the PSD of the output of  $\SSS$. (b) shows the output of $\SSS$ with $v_k \equiv 2$ and $y_k = 500$ for all $k \in [0,6].$ (c) shows the output of $\SM$ with $v_k \equiv 2$ and $y_{\rmm,k} = 0$ for all $k \in [0,6].$ (d) shows the output of $\SSS$ on $[1000,1050]$ and the output of $\SM$ on $[995,1045]$. (e) shows the true and estimated nonlinearities. (f) and (g) show the frequency responses of the linear dynamics of $\SSS$ and $\SM$.}
    \label{ex3_ID_noiseless}
\end{figure}

\begin{figure}[h]
    \centering
    \includegraphics[width=0.7\textwidth]{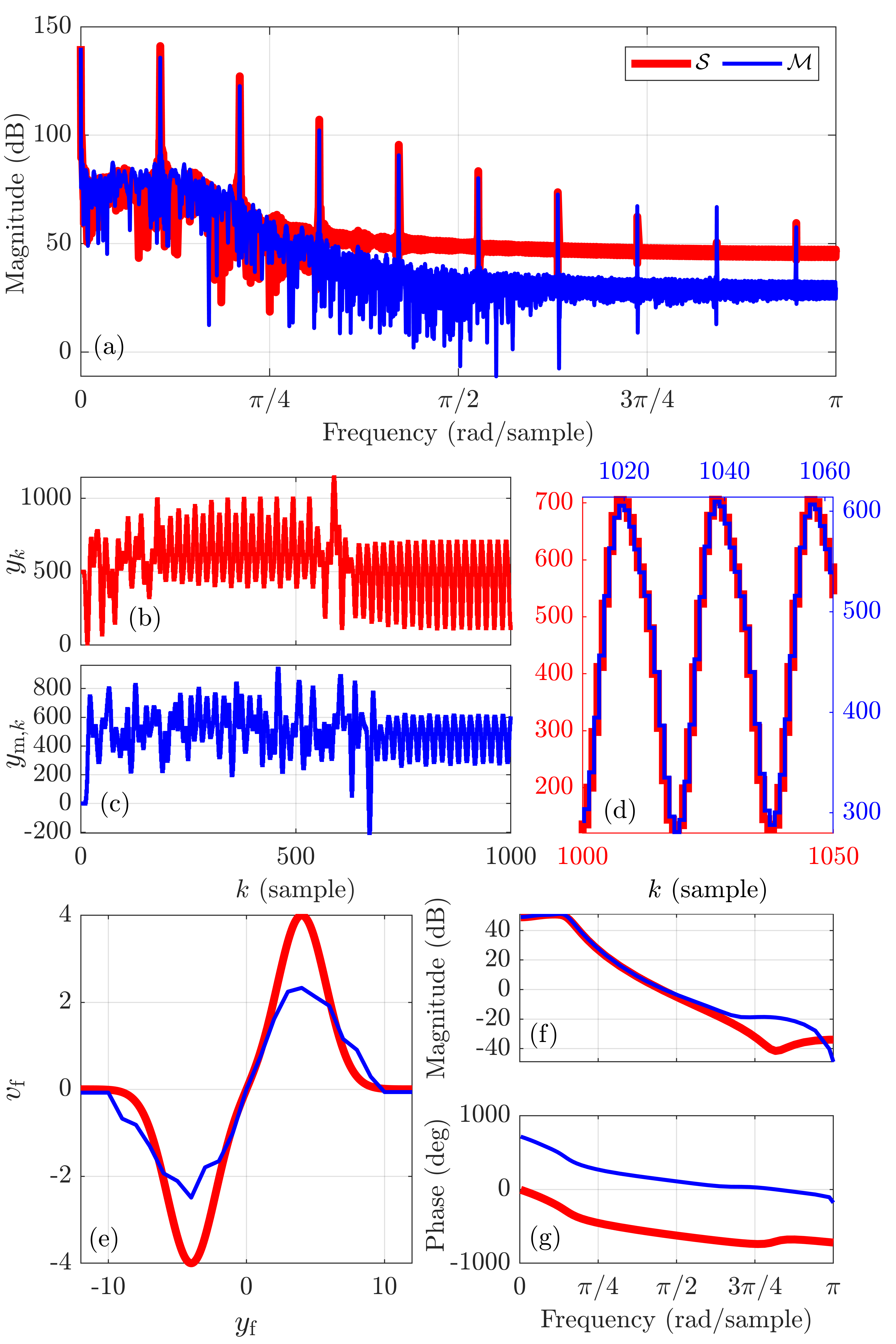}
    \caption{Example \ref{ex3}: Least-squares identification of DTTDL/CPL model parameters using  noisy measurements with $\hat{n} = 10$ and $\hat{d} = d.$ (a) compares the PSD of the output of $\SM$ with the PSD of the output of  $\SSS$. (b) shows the output of $\SSS$ with $v_k \equiv 2$ and $y_k = 500$ for all $k \in [0,6].$ (c) shows the output of $\SM$ with $v_k \equiv 2$ and $y_{\rmm,k} = 0$ for all $k \in [0,6].$ (d) shows the output of $\SSS$ on $[1000,1050]$ and the output of $\SM$ on $[1012,1062]$. (e) shows the true and estimated nonlinearities. (f) and (g) show the frequency responses of the linear dynamics of $\SSS$ and $\SM$.}
    \label{ex3_ID_noise}
\end{figure}

\begin{figure}[h]
    \centering
    \includegraphics[width=0.7\textwidth]{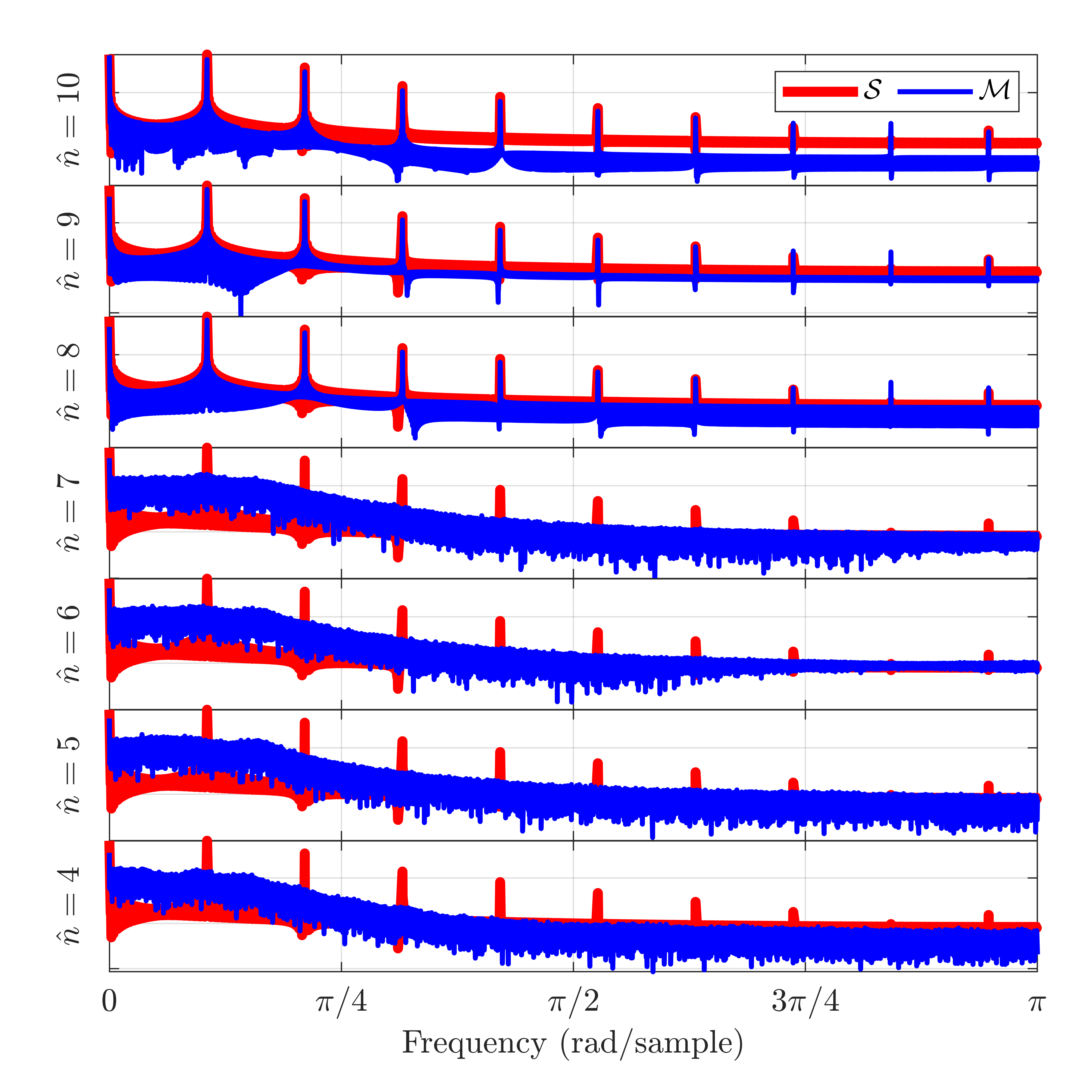}
    \caption{Example \ref{ex3}: For $\hat{n}\in\{4,5,\ldots,9,10\}$ and $\hat{d} = 0,$ these plots compare the PSD of the output of $\SM$ identified using noisy measurements with the PSD of the output of $\SSS$.}
    \label{ex3_ID_grid}
\end{figure}

To obtain data for identification, $y_0, \ldots, y_6$ are generated randomly, and, for all $k\geq 0,$ $v_k$ is a gaussian random variable with mean 3 and standard deviation $\sqrt{5}.$
For all $k\geq 7,$ $y_k$ is generated by simulating $\SSS$ with \eqref{ex3_eq}.
For least-squares identification of the DTTDL/CPL model parameters, we let 
$
    \hat{c} = {[\begin{array}{ccccccc} -10 & -9 & \cdots & 9 & 10 \end{array}]}^\rmT
$
and $\hat{\beta} = \beta,$ and we apply RLS with $\theta_0 = 0,$ $P_0 = 10^6$ and $\lambda = 1$ using data in $[100,100000].$
The standard deviation of the sensor noise is chosen to be $\sqrt{2.5},$ which yields a measurement SNR of approximately $55$ dB.
To assess the accuracy of the identified model, the input $v_k \equiv 2$ is applied to the system $\SSS$ with the initial conditions $y_k = 500$ for all $k\in[0,6],$ as well as the identified model $\SM$ with the initial conditions $y_{\rmm,k} = 0$ for all $k\in[0,6].$
%
% %
The response of the identified model based on  noiseless measurements with $\hat{n} = n$ and $\hat{d} = d$ is shown in Figure \ref{ex3_ID_noiseless}, 
and the response of the identified model based on  noisy measurements with $\hat{n} = 10$ and $\hat{d} = d$ is shown in Figure \ref{ex3_ID_noise}. 
Figure \ref{ex3_ID_grid} compares the PSD of the output of $\SM$ for $\hat{n} \in \{4, 5, \ldots, 9, 10\}$ and $\hat{d} = 0$ obtained using noisy measurements with the PSD of the output of $\SSS$.

\clearpage
\refstepcounter{subsection}
\subsection*{Example \ref{ex4}: CTTDL system with $C^\infty,$ monotonic, odd $\SN$} \label{ex4}

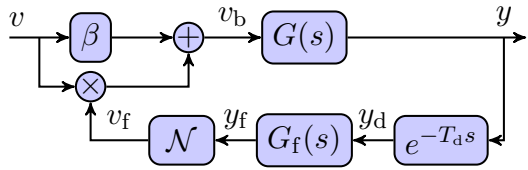
\begin{figure}[h]
    \centering
    \resizebox{0.5\columnwidth}{!}{%
    \begin{tikzpicture}[>={stealth'}, line width = 0.25mm]
    \node [input, name=input] {};
    \node [smallblock, rounded corners, right of=input, minimum height = 0.5cm, minimum width = 0.5cm] (beta) {$\beta$};
    \node [sum, right = 0.75cm of beta] (sum1) {};
    \node[draw = white] at (sum1.center) {$+$};
    \node [smallblock, rounded corners, right = 0.7cm of sum1, minimum height = 0.6cm, minimum width = 0.8cm] (system) {$G(s)$};
    
    \draw [->] (sum1) -- node[name=usys, above] {$v_\rmb$} (system);
    \node [output, right = 2.2cm of system] (output) {};
    \node [smallblock, rounded corners, below = 0.6cm of system, minimum height = 0.6cm, minimum width = 0.8cm](diff){$G_\rmf(s)$};
    \node [smallblock, rounded corners, right = 0.5cm of diff, minimum height = 0.6cm, minimum width = 0.8cm] (delay) {$e^{-T_\rmd s}$};
    \node [smallblock, rounded corners, left = 0.5cm of diff, minimum height = 0.6cm, minimum width = 0.8cm](satq){$\SN$};
    \node [mult, below = 0.1cm of beta, minimum size=0.35cm] (mult1) {};
    \node [draw = white] at (mult1.center) {$\times$};
    
    \draw [draw,->] (input) -- node [name=u]{} node [very near start, above] {$v$} (beta);
    \draw [->] (u.center) |- (mult1);
    \draw [->] (beta) -- (sum1);
    \draw [->] (mult1) -| (sum1);
    \draw [->] (satq) -|  
    node [near start, above] {$v_\rmf$} (mult1);
    \draw [->] (system) -- node [name=y, very near end]{} node [very near end, above] {$y$}(output);
    \draw [->] (y.center) |- (delay);
    \draw [->] (delay) -- node [above] {$y_\text{d}$} (diff);
    \draw [->] (diff) -- node [above] {$y_{\text{f}}$}(satq);
    \end{tikzpicture}
    }
    \caption{\footnotesize Continuous-time, time-delayed Lur'e  model with constant input $u$ and bias-generation mechanism.}
    \label{CT_TDL_offset_blk_diag}
\end{figure}

Let  $T_\rmd \geq 0$ be the  delay time, and let $t \geq 0.$  
Furthermore, let $G$ be an $n$th-order strictly proper SISO transfer function with minimal realization $(A,B,C)$ and  state $x(t) \in \BBR^n.$
Finally, let $G_\rmf$ be a SISO washout filter with realization $(A_{\rmf},B_{\rmf},C_{\rmf}, D_{\rmf})$ and  state $x_\rmf(t)\in\BBR.$ Then, for all $t\ge T_\rmd,$ the closed-loop dynamics of the continuous-time, time-delayed Lur'e (CTTDL) model shown in Figure \ref{CT_TDL_offset_blk_diag} are given by
\begin{equation}
    \begin{bmatrix}  \dot{x} (t)\\ {\dot{x}}_{\rmf} (t) \end{bmatrix}
    = \begin{bmatrix} A & 0 \\ 0 & A_{\rmf} \end{bmatrix}
    \begin{bmatrix} x(t)\\ x_{\rmf} (t)  \end{bmatrix}
    + \begin{bmatrix} 0 & 0\\ B_{\rmf} C & 0  \end{bmatrix}
    \begin{bmatrix} x(t - T_{\rmd})\\ x_{\rmf} (t - T_{\rmd})  \end{bmatrix}
     + \begin{bmatrix} B \\ 0 \end{bmatrix} v_{\rm b}(t), \label{TDLeqn}
\end{equation}
with output
\begin{align}
    y (t) = \matl{cc} C & 0 \matr
    \matl{c} x (t)\\ x_{\rmf} (t)\matr ,
    \label{TDLeqnlin}
\end{align}
the bias-generation mechanism
\begin{align}
    v_{\rm b} (t) = v(t) (\beta + v_{\rm f} (t)),
  \label{TDLbias}
\end{align}
and  signals
\begin{align}
    v_{\rm f} (t) &= \SN(y_{\rmf} (t)), \\
    y_{\rmf} (t) &= C_{\rmf} x_{\rmf}(t) + D_{\rmf} y_{\rmd}(t), \\
    y_{\rm d} (t) &= y (t - T_{\rmd}). \label{TDLeqnfb}
\end{align}
The resulting bias $\bar{y}$ of the periodic response is thus given by $\bar{y} = v \beta G(0).$
Note that the initialization of \eqref{TDLeqn} depends on $x_\rmf(T_\rmd)$ as well as $x(t)$ for all $t\in[0,T_\rmd]$.
To compute the solution of the delay differential equations (DDEs) \eqref{TDLeqnlin}--\eqref{TDLeqnfb}, a Runge-Kutta DDE method is used with an interpolant to approximate the delayed terms, as in \cite[pp. 156--158]{bellen2003}.
In this paper, 4th-order Runge-Kutta is used with a linear interpolant and a fixed time step of 0.001 s.

Consider the CTTDL system $\SSS$ with $\beta = 50,$ $T_\rmd = 0.1$ s,  $G_{\rmf} (\bfp) = \bfp / (\tau\bfp + 1)$ with realization $A_\rmf = -1/\tau,$ $B_\rmf = 1,$ $C_\rmf = -1/\tau^2,$ $D_\rmf = 1/\tau,$ where $\tau = 0.001$ s, 
\begin{equation}
    G(\bfp) = \frac{\bfp + 2.5}{\bfp^2 + \bfp + 6.5},
\end{equation}
with realization
\begin{equation}
    A = \begin{bmatrix} -1 & -6.5 \\ 1 & 0 \end{bmatrix}, \hspace{1em} B = \begin{bmatrix} 1 \\  0 \end{bmatrix}, \hspace{1em} C = \begin{bmatrix} 1 & 2.5 \end{bmatrix},
\end{equation}
and $\SN (y_\rmf) = 5 \tanh (y_\rmf / 5).$

To obtain data for identification, let $y_0 \in \BBR$ and $y(t) = y_0$ for all $t\in[0,T_\rmd].$ 
Therefore,  for all $t\in[0,T_\rmd],$ $\dot{y}(t) = 0$, and thus $x(t) = x_0$, where
\begin{equation}
    x_0 = {\matl{c} C\\ CA\matr}^{-1} \matl{c} y_0\\ 0\matr.
\end{equation}
Furthermore, since $y_\rmd (T_\rmd) = y_0,$ let $x_\rmf (T_\rmd) = \tau y_0 = 0.001 y_0,$ such that $\dot{x}_{\rmf} (T_\rmd) = y_\rmf (T_\rmd) = 0.$
For all $t>T_\rmd,$ $y(t)$ is computed using the DDE method described above, with $y_0 = 0.$ 
For all $k \geq 0,$  sampled output is $y_k = y(kT_\rms)$, where $T_\rms = 0.1$ s is the sample time.

\begin{figure}[h]
    \centering
    \includegraphics[width=0.7\textwidth]{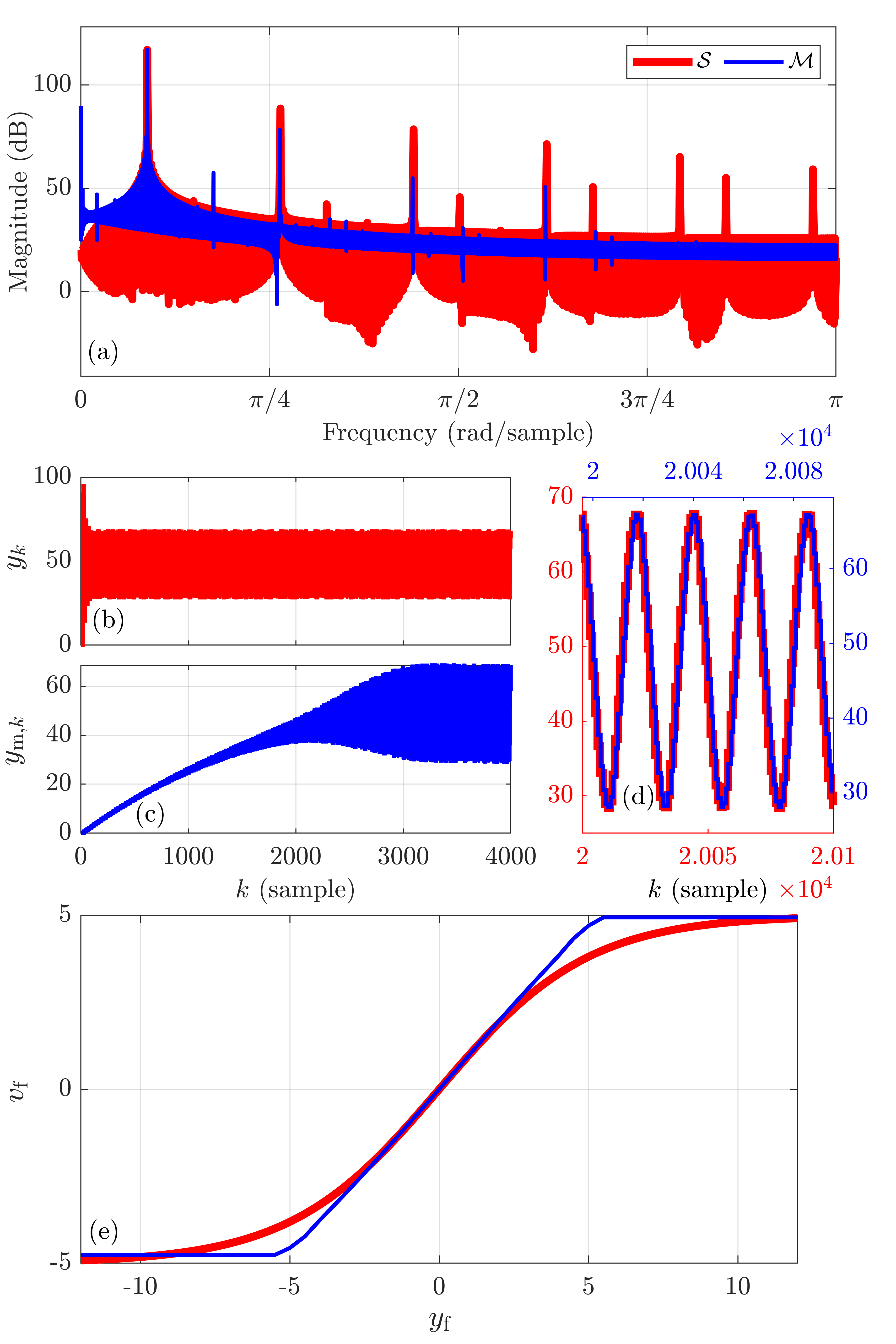}
    \caption{Example \ref{ex4}: Least-squares identification of DTTDL/CPL model parameters for constant input measurements using noiseless measurements arising from a constant input $v,$ with $\hat{n} = 12,$ $\hat{d} = 5,$ and $\beta_{\rm LS} = 5.$ (a) compares the PSD of the output of $\SM$ with the PSD of the output of  $\SSS$. (b) shows the sampled output of $\SSS$ with $v(t) \equiv 5$ and $y_k = 0$ for all $k \in [0, 1].$ (c) shows the output of $\SM$ with $v_k \equiv 5$ and $y_{\rmm,k} = 0$ for all $k \in [0,17].$ (d) shows the sampled output of $\SSS$ on $[20000,20100]$ and the output of $\SM$ on $[19994,20094]$. (e) shows the true and estimated nonlinearities.}
    \label{ex4_ID_const_noiseless}
\end{figure}

In the first case, the data are obtained by subjecting $\SSS$ to the constant input $v(t) \equiv 2.5$.
For least-squares identification of the DTTDL/CPL model parameters with constant input, we let $\hat{n} = 12,$ $\hat{d} = 5,$ $\beta_{\rm LS} = 5,$ and
$
    \hat{c} = {[\begin{array}{ccccccc} -6 & -5.5 & \cdots & 5.5 & 6 \end{array}]}^\rmT,
$
and we apply RLS with $\theta_0 = 0,$ $P_0 = {10}^{2}$ and $\lambda = 1$ using data in $[100, 100000].$
To assess the accuracy of the identified model, $v(t) \equiv 5$ is applied to $\SSS$ with the initial conditions $y(0) = 0,$ $x(t) = 0$ for all $t\in[0,T_\rmd]$ and $x_\rmf (T_\rmd) = 0,$ and $v_k \equiv 5$ is applied to the identified model $\SM$ with the initial conditions $y_{\rmm,k} = 0$ for all $k\in[0,17].$
The response of the identified model based on noiseless measurements is shown in Figure \ref{ex4_ID_const_noiseless}.

\begin{figure}[h]
    \centering
    \includegraphics[width=0.7\textwidth]{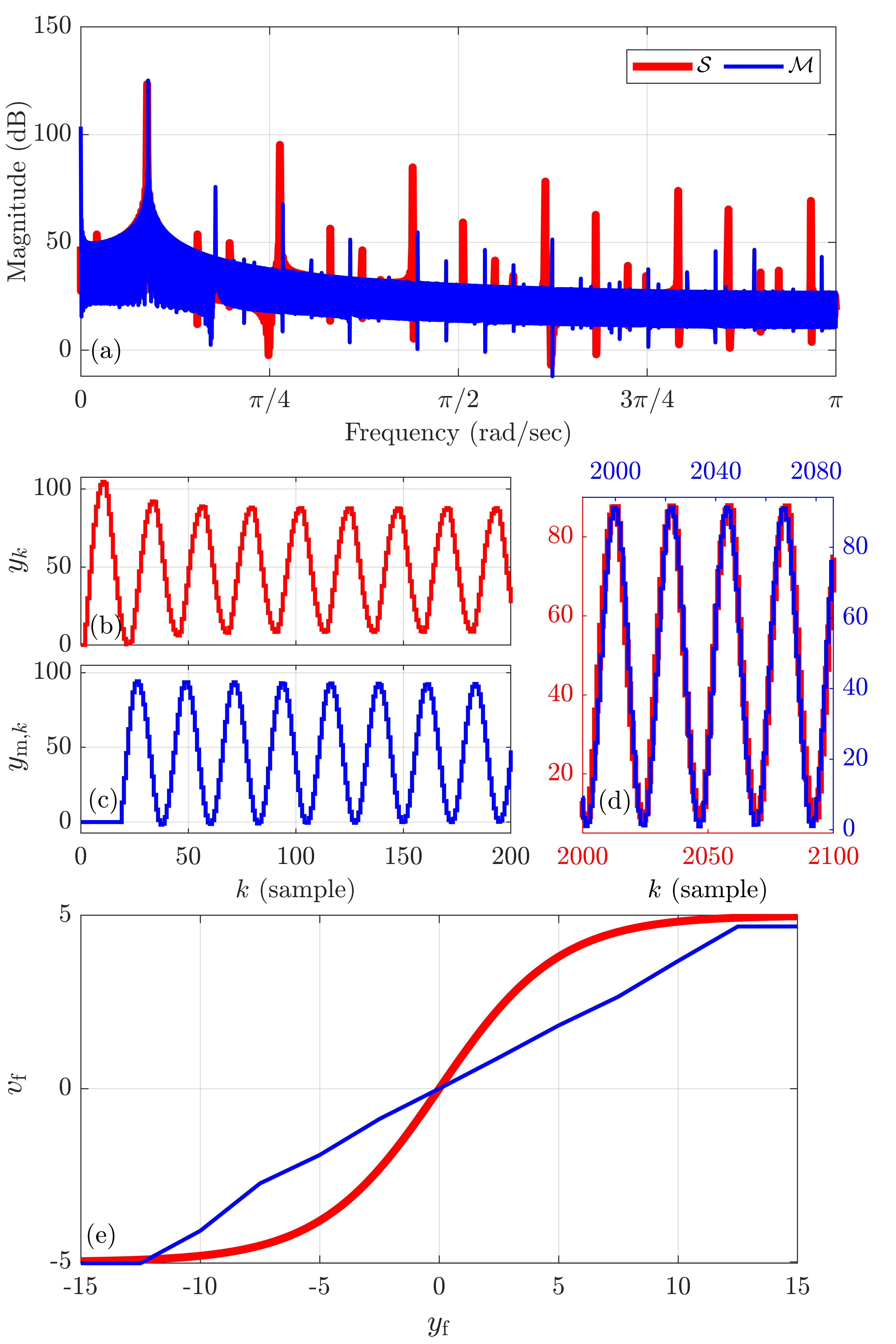}
    \caption{Example \ref{ex4}: Least-squares identification of DTTDL/CPL model parameters using noiseless measurements arising from a non-constant input $v,$ with $\hat{n} = 12,$ $\hat{d} = 4,$ and $\hat{\beta} = 25$. (a) compares the PSD of the output of $\SM$ with the PSD of the output of  $\SSS$. (b) shows the sampled output of $\SSS$ with $v(t) \equiv 5$ and $y_k = 0$ for all $k \in {0, 1}.$  (c) shows the output of $\SM$ with $v_k \equiv 5$ and $y_{\rmm,k} = 0$ for all $k \in [0,16].$ (d) shows the sampled output of $\SSS$ on $[2000,2100]$ and the output of $\SM$ on $[1988,2088]$. (e) shows the true and estimated nonlinearities.}
    \label{ex4_ID_wn_noiseless}
\end{figure}

\begin{figure}[h]
    \centering
    \includegraphics[width=0.7\textwidth]{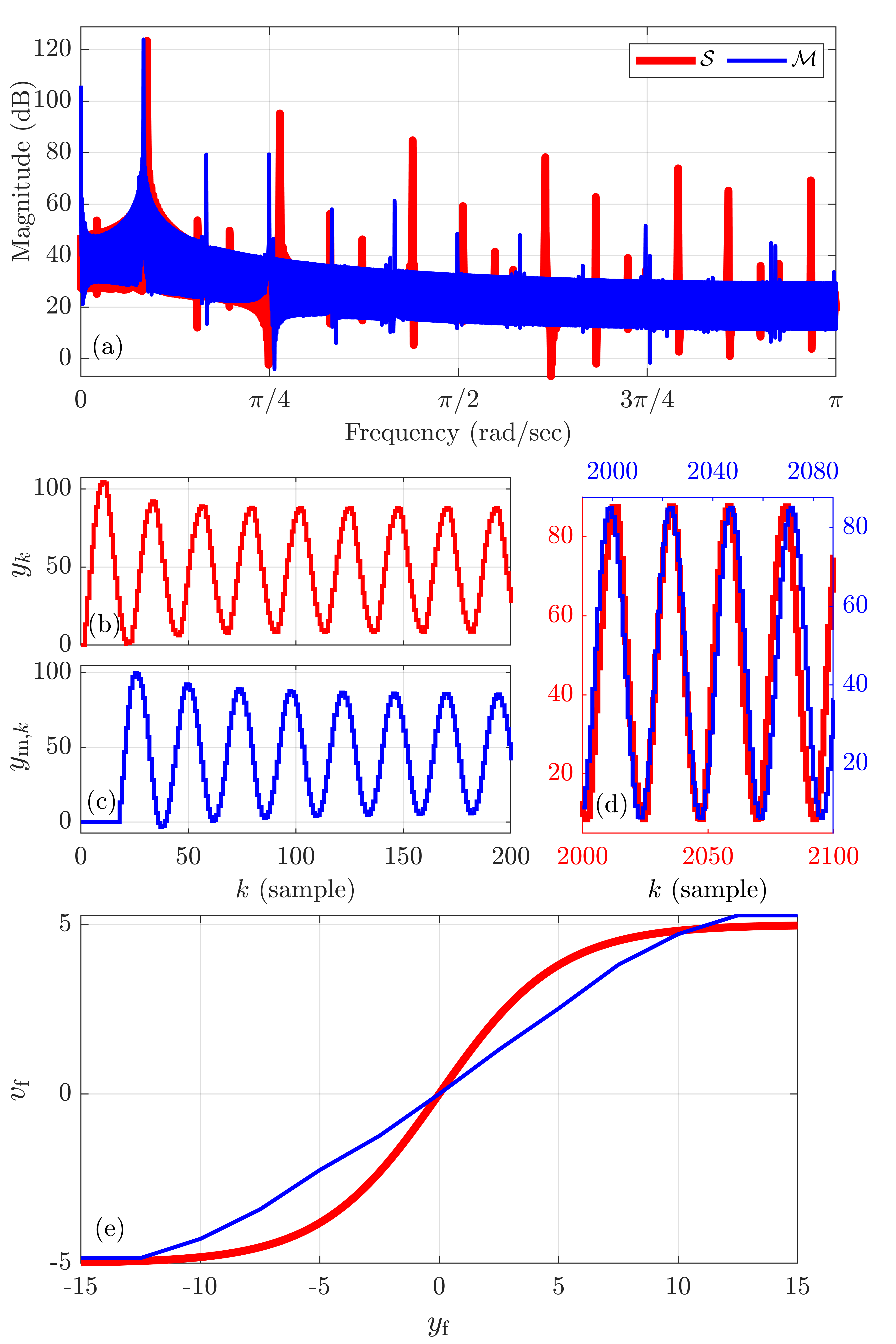}
    \caption{Example \ref{ex4}: Least-squares identification of DTTDL/CPL model parameters using noisy measurements arising from a non-constant input $v,$ with $\hat{n} = 12,$ $\hat{d} = 4,$ and $\hat{\beta} = 15.$ (a) compares the PSD of the output of $\SM$ with the PSD of the output of  $\SSS$. (b) shows the sampled output of $\SSS$ with $v(t) \equiv 5$ and $y_k = 0$ for all $k \in {0, 1}.$ (c) shows the output of $\SM$ with $v_k \equiv 5$ and $y_{\rmm,k} = 0$ for all $k \in [0,16].$ (d) shows the sampled output of $\SSS$ on $[2000,2100]$ and the output of $\SM$ on $[1989,2089]$. (e) shows the true and estimated nonlinearities.}
    \label{ex4_ID_wn_noisy}
\end{figure}

In the second case, the data are obtained by subjecting  $\SSS$ to a piecewise-constant input such that, for all $k\ge0,$ it follows that, for all $t\in[k T_\rms, (k+1)T_\rms),$ $v(t) = \omega_k,$ where $\omega_k$ is a gaussian random variable with mean $2.5$ and standard deviation $\sqrt{0.5}.$
For least-squares identification of the DTTDL/CPL model parameters, we let $\hat{n} = 12,$ $\hat{d} = 4,$ and
$
    \hat{c} = {[\begin{array}{ccccccc} -12.5 & -10 & \cdots & 10 & 12.5 \end{array}]}^\rmT,
$
and we apply RLS with $\theta_0 = 0,$ $P_0 = {10}^{2}$ and $\lambda = 1$ using data in $[100, 100000].$
The standard deviation of the sensor noise is chosen to be $\sqrt{2},$ which yields a measurement SNR of approximately 30 dB.
To assess the accuracy of the identified model, $v(t) \equiv 5$ is applied to $\SSS$ with the initial conditions $y(0) = 0,$ $x(t) = 0$ for all $t\in[0,T_\rmd]$, and $x_\rmf (T_\rmd) = 0.$ 
The sampled input $v_k \equiv 5$ is applied to the identified model $\SM$ with the initial conditions $y_{\rmm,k} = 0$ for all $k\in[0,16].$
The response of the identified model based on noiseless measurements with $\hat{\beta} = 25$ is shown in Figure \ref{ex4_ID_wn_noiseless}.
The response of the identified model based on noisy measurements with $\hat{\beta} = 15$ is shown in Figure \ref{ex4_ID_wn_noisy}.
\clearpage
\refstepcounter{subsection}
\subsection*{Example \ref{ex5}: Van der Pol system with bias} \label{ex5}
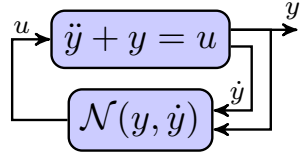
\begin{figure}[h!]
    \centering
    \resizebox{0.3\columnwidth}{!}{%
    \begin{tikzpicture}[>={stealth'}, line width = 0.25mm]
        \node [smallblock,rounded corners,fill=blue!20, minimum height = 0.6cm , minimum width = 0.6cm] (LinDyn){$\ddot{y} + y = u$};
        
        \node [smallblock,rounded corners,fill=blue!20, minimum height = 0.6cm , minimum width = 0.6cm, below = 0.2cm of LinDyn] (NLBlk){$\mathcal{N}(y,\dot{y})$};
        
        \draw[->] (NLBlk.west)-|([xshift=-0.4cm]LinDyn.west)--node [above, xshift = -0.1cm, yshift = -0.075cm]{\scriptsize$u$}(LinDyn.west);
        
        \draw[->] ([yshift=0.1cm]LinDyn.east)-|([xshift=0.6cm,yshift=-0.1cm]NLBlk.east)--([yshift=-0.1cm]NLBlk.east);
        
        \draw[->] ([yshift=0.1cm]LinDyn.east)--node [above, xshift = 0.3cm, yshift = -0.025cm]{\scriptsize$y$}([xshift=0.7cm,yshift=0.1cm]LinDyn.east);
        
        \draw[->] ([yshift=-0.075cm]LinDyn.east)-|([xshift=0.4cm,yshift=0.1cm]NLBlk.east)--node[above, xshift = 0.05cm, yshift = -0.075cm]{\scriptsize$\dot{y}$}([yshift=0.1cm]NLBlk.east);
    \end{tikzpicture}
    }
    \caption{Block representation of the Van der Pol system, where $\SN(y,\dot{y}) = \mu_0(1 - y^2)\dot{y}.$}
    \label{VdP_blk}
\end{figure}

Let the $\SSS$ be the continuous-time Van der Pol system
\begin{equation}
    \Ddot{y} + \mu_0(y^2-1)\dot{y} + y = 0,
\end{equation}
where $\mu_0$ is a constant parameter.
Figure \ref{VdP_blk} represents $\SSS$ as a Lur'e system.

To obtain data for identification, let $\mu_0 = 1,$ $y(0) = 0.1,$ and $\dot{y}(0) = 0.$ For all $t>0,$ the Van der Pol system is simulated using ode45, and the output is sampled with sample time $T_\rms = 0.1$ s.
The integration accuracy of ode45 is set so that approximately 160 integration steps are implemented within each sample interval.
A bias $\bar{y}$ is added to all sampled measurements so that, for all $k \geq 0,$ the biased output is $y_k = y(kT_\rms) + \bar{y}$, where $\bar{y} = 10$.
Finally, for identification purposes, it is assumed that $v(t) \equiv 1$ is applied to $\SSS.$

\begin{figure}[h]
    \centering
    \includegraphics[width=0.7\textwidth]{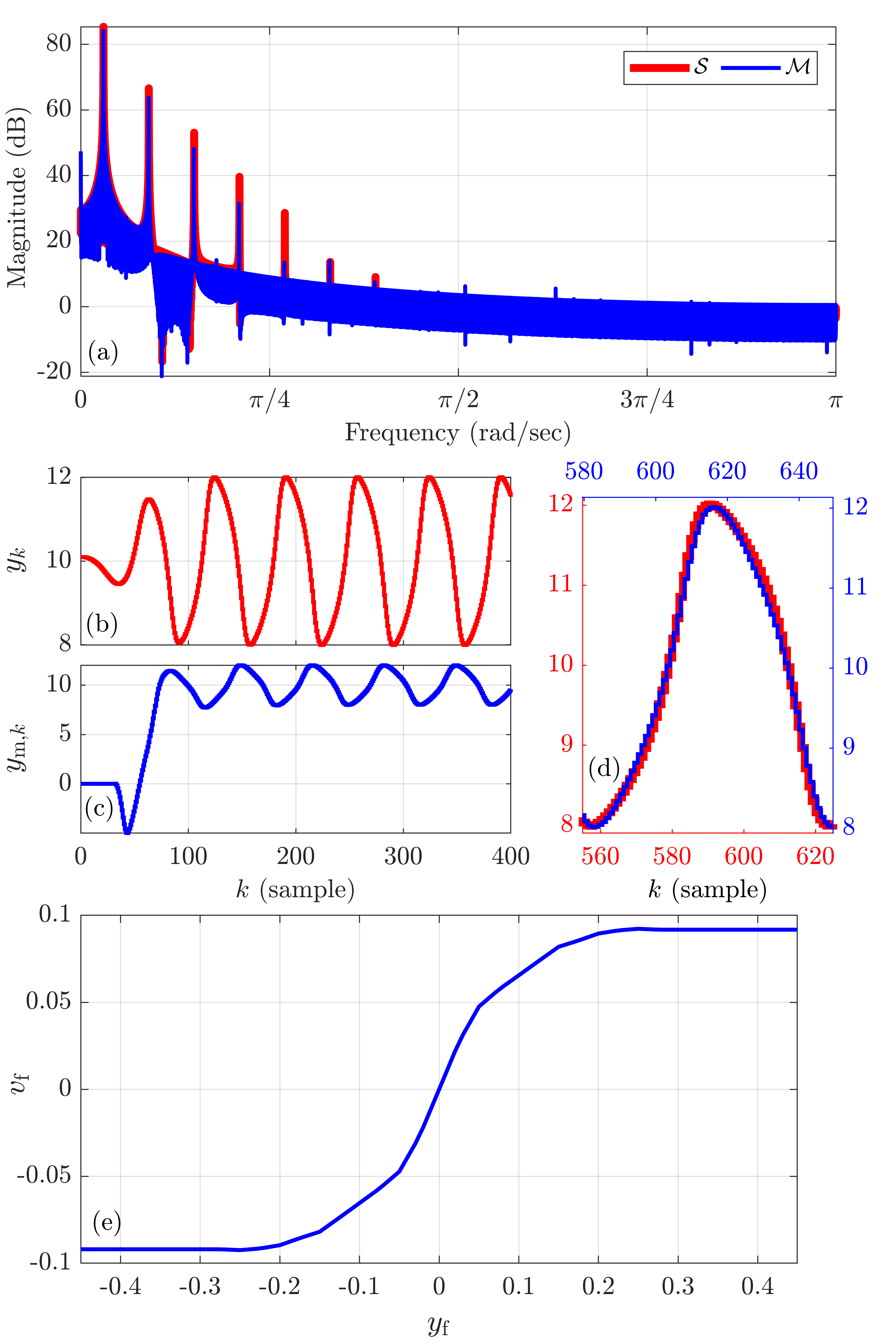}
    \caption{Example \ref{ex5}: Least-squares identification of DTTDL/CPL model parameters for constant input measurements using  noiseless measurements with $\hat{n} = 12,$ $\hat{d} = 19,$ and $\beta_{\rm LS} = -5.$ (a) compares the PSD of the output of $\SM$ with the PSD of the output of  $\SSS$. (b) shows the biased sampled output of $\SSS.$ (c) shows the output of the $\SM$ with $v_k \equiv 1$ and $y_{\rmm,k} = 0$ for all $k \leq 31$. (d) shows the sampled output of $\SSS$ on $[555,625]$ and the output of $\SM$ on $[582,652].$ (e) shows the estimated nonlinear feedback mapping.}
    \label{ex5_ID_mu_1}
\end{figure}

For least-squares identification of the DTTDL/CPL model parameters with constant input, we let 
$
    \hat{c} = {[\begin{array}{ccccc} -0.3 & -0.275 & \cdots & 0.275 & 0.3 \end{array}]}^\rmT,
$
$\hat{n} = 12$ and $\hat{d} = 19,$ and $\beta_{\rm LS} = -5,$
and we apply RLS with $\theta_0 = 0,$ $P_0 = {10}^{2}$ and $\lambda = 1$ using data in $[225, 20000].$
To assess the accuracy of the identified model, $v_k \equiv 1$ is applied to the identified model $\SM$ with the initial conditions $y_{\rmm,k} = 0$ for all $k\in[0,31].$
The response of the identified model based on noiseless measurements is shown in Figure \ref{ex5_ID_mu_1}.

\begin{figure}[h]
    \centering
    \includegraphics[width=0.7\textwidth]{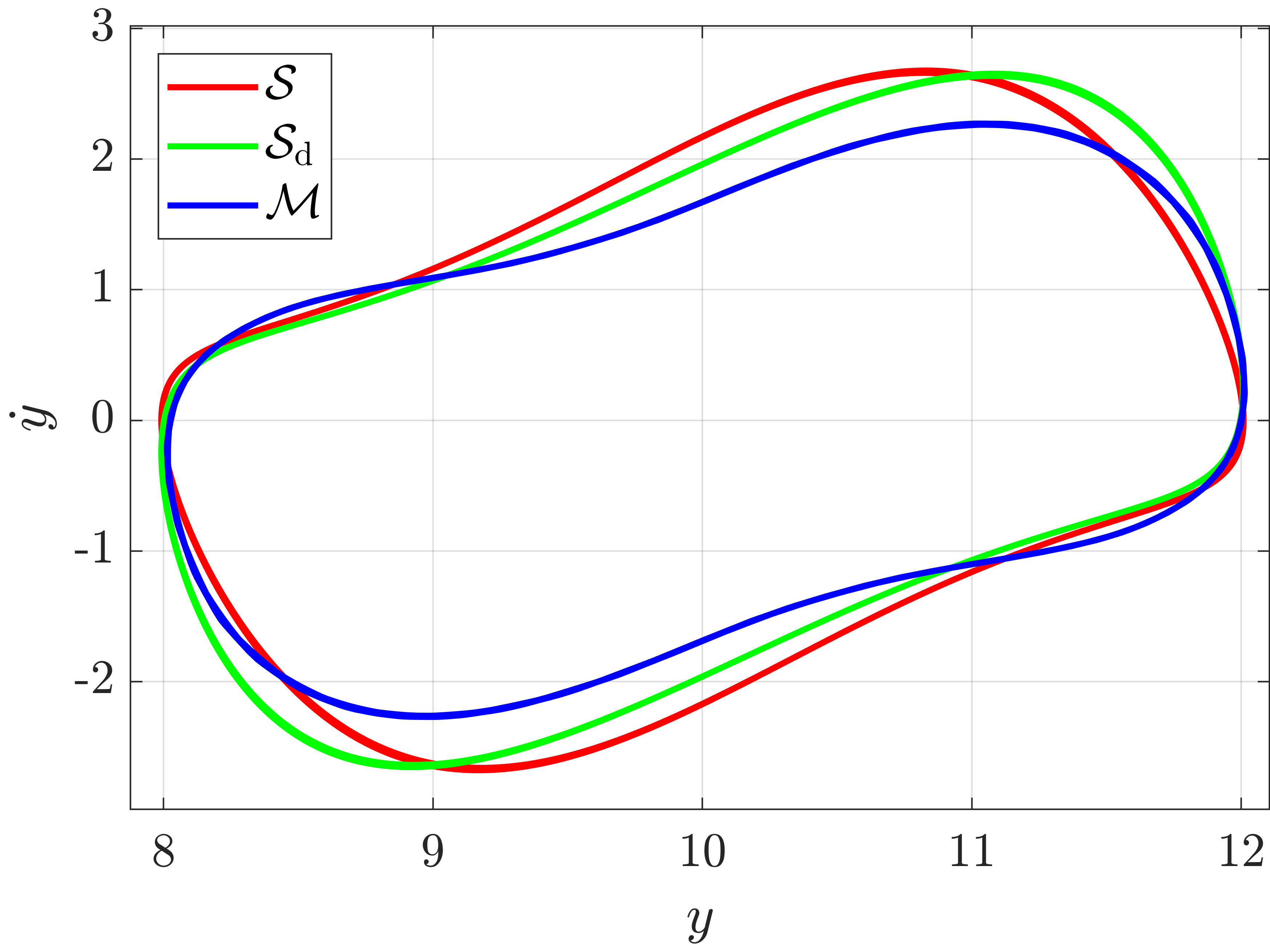}
    \caption{Example \ref{ex5}: Phase portraits of the response of the continuous-time Van der Pol system $\SSS$ with $\mu_0 = 1,$ the response of the discrete-time system $\SSS_\rmd,$ whose output is the sampled output of $\SSS,$ and the response of the identified model $\SM.$ The derivative of the output of $\SSS_\rmd$ and $\SM$ is approximated by using  $\dot{y}_k = \tfrac{y_{k+1} - y_{k-1}}{2 T_\rms}.$}
    \label{ex5_ID_mu_1_phase}
\end{figure}

Let $\SSS_\rmd$ be the system whose output is the sampled output of $\SSS$ and where the derivative of the sampled output is approximated by $\dot{y}_k = \tfrac{y_{k+1} - y_{k-1}}{2 T_\rms}.$
Figure \ref{ex5_ID_mu_1_phase} compares the phase portraits of the continuous-time system $\SSS,$ the discrete-time system $\SSS_\rmd,$ and the identified model $\SM$ using  $\dot{y}_k = \tfrac{y_{k+1} - y_{k-1}}{2 T_\rms}$ to approximate the derivative of the output.

\clearpage

\refstepcounter{subsection}
\subsection*{Example \ref{ex6}: Predator-prey Lotka-Volterra system} \label{ex6}
Let the $\SSS$ be the continuous-time predator-prey Lotka-Volterra system
\begin{align}
    \dot{y} &=  \zeta y - \varrho xy,\\
    \dot{x} &= -\xi x + \varphi xy,
\end{align}
where $\zeta, \varrho, \xi,\varphi$ are constant parameters.

To obtain data for identification, let $\zeta = 2/3,$ $\varrho = 4/3,$ $\xi = 1,$ $\varphi = 1,$ and $y(0) = x(0) = 1.$ For all $t>0,$ the Lotka-Volterra system is simulated using ode45, and the output is sampled with sample time $T_\rms = 0.1$ s.
The integration accuracy of ode45 is set so that approximately 160 integration steps are implemented within each sample interval.
Finally, for identification purposes, it is assumed that $v(t) \equiv 1$ is applied to $\SSS.$

\begin{figure}[h]
    \centering
    \includegraphics[width=0.7\textwidth]{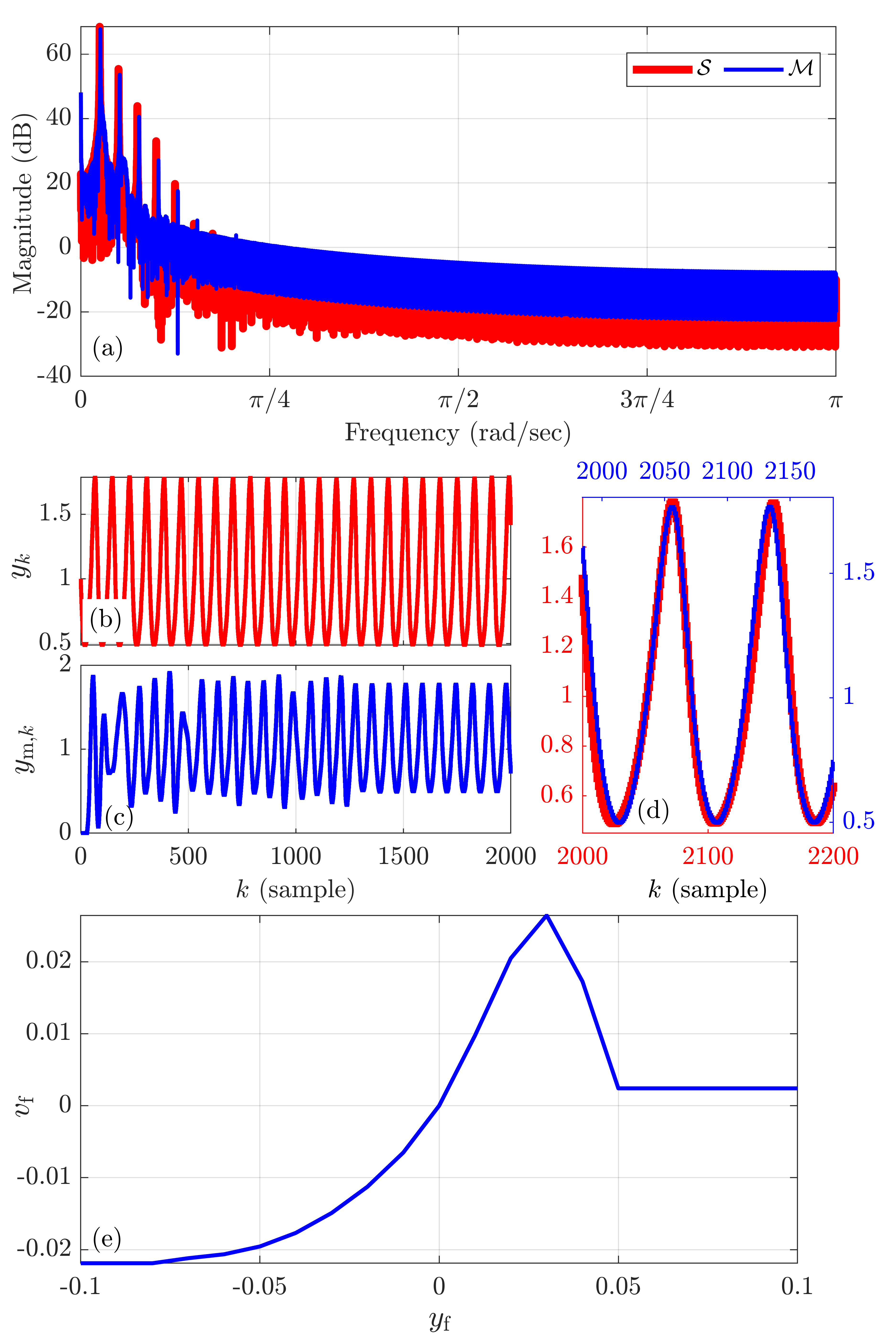}
    \caption{Example \ref{ex6}: Least-squares identification of DTTDL/CPL model parameters for constant input measurements using  noiseless measurements with $\hat{n} = 12,$ $\hat{d} = 13,$ and $\beta_{\rm LS} = -5.$ (a) compares the PSD of the output of $\SM$ with the PSD of the output of  $\SSS$. (b) shows the biased sampled output of $\SSS.$ (c) shows the output of the $\SM$ with $v_k \equiv 1$ and $y_{\rmm,k} = 0$ for all $k \leq 25$. (d) shows the sampled output of $\SSS$ on $[2000,2200]$ and the output of $\SM$ on $[1985,2185].$ (e) shows the estimated nonlinear feedback mapping.}
    \label{ex6_ID}
\end{figure}

For least-squares identification of the DTTDL/CPL model parameters with constant input, we let 
$
    \hat{c} = {[\begin{array}{ccccc} -0.08 & -0.07 & \cdots & 0.05 & 0.06 \end{array}]}^\rmT,
$
$\hat{n} = 12$, $\hat{d} = 13,$ and $\beta_{\rm LS} = -5,$
and we apply RLS with $\theta_0 = 0,$ $P_0 = {10}^{2},$ and $\lambda = 1$ using data in $[500, 10000].$
To assess the accuracy of the identified model, $v_k \equiv 1$ is applied to the identified model $\SM$ with the initial conditions $y_{\rmm,k} = 0$ for all $k\in[0,25].$
The response of the identified model based on noiseless measurements is shown in Figure \ref{ex6_ID}.

\begin{figure}[h]
    \centering
    \includegraphics[width=0.7\textwidth]{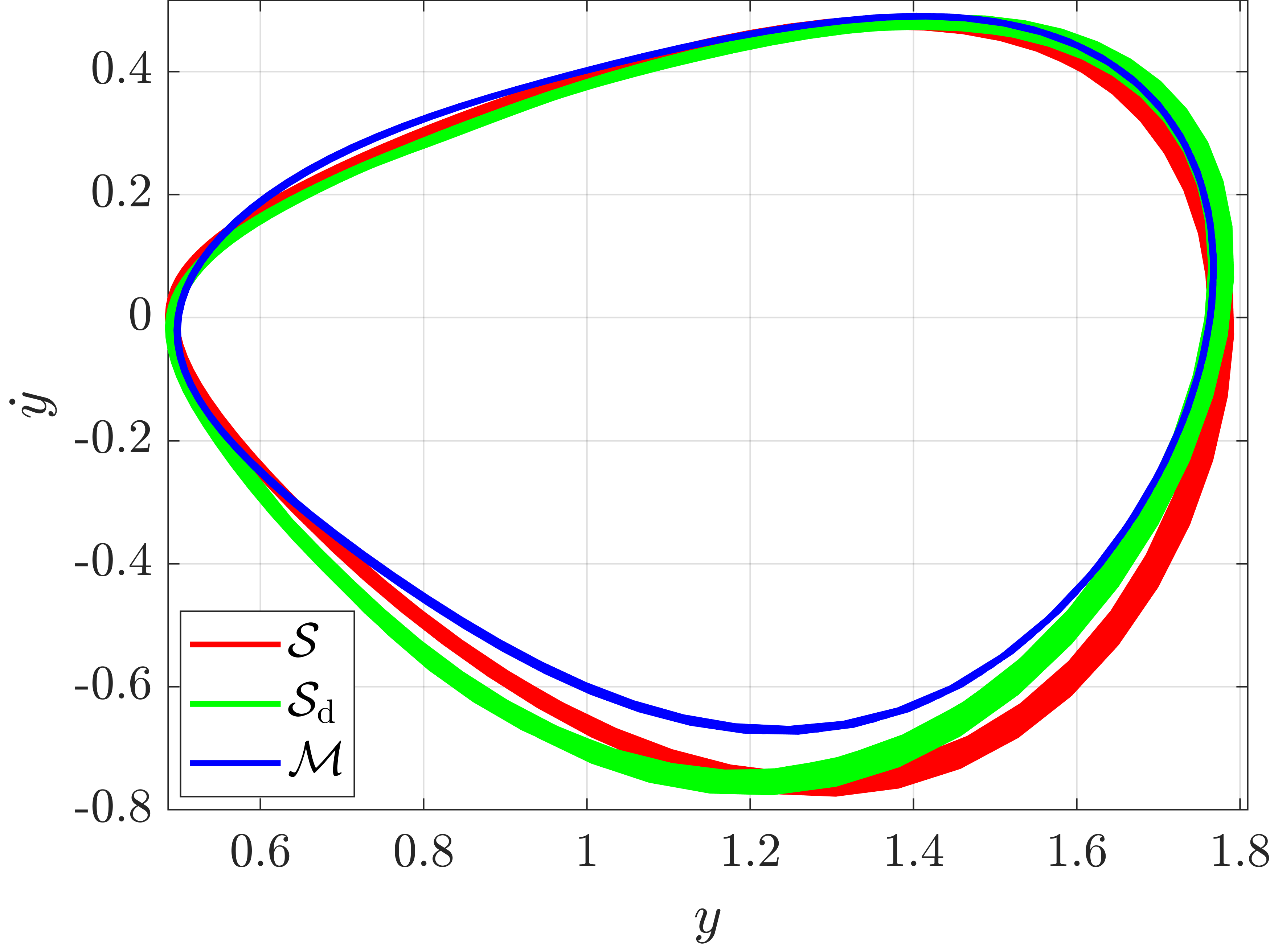}
    \caption{Example \ref{ex6}: Phase portraits of the response of the continuous-time Lotka-Volterra system $\SSS$ with $\zeta = 2/3, \varrho = 4/3, \xi = 1, \varphi = 1,$ and $y(0) = x(0) = 1,$ the response of the discrete-time system $\SSS_\rmd,$ whose output is the sampled output of $\SSS,$ and the response of the identified model $\SM.$ The derivative of the output of $\SSS_\rmd$ and $\SM$ is approximated by using  $\dot{y}_k = \tfrac{y_{k+1} - y_{k-1}}{2 T_\rms}.$}
    \label{ex6_ID_phase}
\end{figure}

Let $\SSS_\rmd$ be the system whose output is the sampled output of $\SSS$ and where the derivative of the sampled output is approximated by $\dot{y}_k = \tfrac{y_{k+1} - y_{k-1}}{2 T_\rms}.$
Figure \ref{ex6_ID_phase} compares the phase portraits of the continuous-time system $\SSS,$ the discrete-time system $\SSS_\rmd,$ and the identified model $\SM$ using  $\dot{y}_k = \tfrac{y_{k+1} - y_{k-1}}{2 T_\rms}$ to approximate the derivative of the output.

\clearpage

\section{Conclusions and future work}

This paper developed a technique for  identification of self-excited systems (SES) based on a discrete-time, time-delayed Lur'e (DTTDL) model.
The nonlinear feedback mapping was chosen to be a continuous, piecewise-linear (CPL) function characterized by its slope in each interval of a user-chosen partition of the real line.
By minimizing a bound on a nonquadratic cost function, linear least-squares techniques were used for parameter estimation within DTTDL/CPL.

Numerical examples included both discrete-time and  continuous-time systems with sampled data.
Of particular interest was the ability of the DTTDL model to reproduce the limit-cycle response of the Van der Pol oscillator and the Lotka-Volterra model.
Although neither of these systems have the structure of a DTTDL model, the system identification technique was able to approximately reproduce the phase-plane dynamics of both systems.

Future research will focus on efficient techniques for determining the user-chosen partition of the real line needed to parameterize the static nonlinear feedback mapping.
Finally, the numerical results motivate a fundamental research question, namely, to what extent can DTTDL/CPL models approximate the response of an arbitrary SES system.

\section{Acknowledgments}

This research was supported by NSF grant CMMI 1634709,
``A Diagnostic Modeling Methodology for
Dual Retrospective Cost Adaptive Control of Complex Systems.''

\bibliographystyle{apacite}
\bibliography{bibliography}

\end{document}